%% file: bflows.tex
\documentclass[titlepage,12pt]{utarticle}   

\usepackage{graphics}
\usepackage{latexsym,amsmath,amssymb,amscd,cite}
\usepackage{epsf}
\usepackage[all]{xy}

\DeclareFontFamily{U}{rsf}{}
\DeclareFontShape{U}{rsf}{m}{n}{<5> <6> rsfs5 <7> <8> <9> rsfs7 <10-> rsfs10}{}
\DeclareMathAlphabet\Scr{U}{rsf}{m}{n}

\def\Z{{\Bbb Z}}
\def\C{{\Bbb C}}
\def\rk{{\rm rk}}
\def\Hom{{\rm Hom}}
\def\End{{\rm End}}
\def\Aut{{\rm Aut}}

\def\DG{{\rm DG}}
\def\DB{{\rm DB}}
\def\P{{\rm P}}

\newcommand{\beq}{\begin{equation}}
\newcommand{\eeq}{\end{equation}}
\newcommand{\bea}{\end{eqnarray}}
\newcommand{\eea}{\end{eqnarray}}
\newcommand{\nn}{\nonumber}

\newcommand{\beql}[1]{\begin{eqnarray}\label{#1}}

\newcommand{\eeql}{\end{eqnarray}}

\long\def\begdel#1\enddel{}
\def\rf#1{(\ref{#1})}

\newcommand{\Tr}{{\rm Tr}}
\newcommand{\diag}{{\rm diag}}

\def\lab#1{^{(#1)}}
\def\Lab#1{^{[#1]}}

\def\half{{1\over2}}

\def\non{\nonumber}

\def\del{\partial}
\def\cR{{\mathcal R}}

\def\BC{{\mathbb C}}

\def\BZ{{\mathbb Z}}

\def\cZ{{\cal Z}}
\def\al{\alpha}
\def\bfone{{\bf 1}}

\def\weff{{\mathcal W}_{\rm eff}}

\def\wlg{W}

\def\hll{\half(\ell_1+\ell_2)}
\def\vp{x}
\def\del{\partial}

\def\cM{{\mathbb M}}
\def\cN{{\mathbb N}}
\def\cP{{\mathbb P}}
\def\Mp{{M_+}}
\def\Mm{{M_-}}
\def\Np{{N_+}}
\def\Nm{{N_-}}

\begin{document}        
\preprint{
  CERN--PH--TH/2004-088\\
  MAD-TH-04-5\\
  {\tt hep-th/0405138}\\
}
\title{D-brane effective action and tachyon condensation
 in topological minimal models}
\author{
Manfred Herbst${}^a$, Calin-Iuliu Lazaroiu${}^b$, Wolfgang Lerche${}^a$
} 
    \oneaddress{
     ${}^a$ Department of Physics, CERN\\
     Theory Division\\
     CH-1211 Geneva 23\\
     Switzerland\\
     \email{firstname.lastname@cern.ch}\\{~}\\
     ${}^b$ 5243 Chamberlin Hall\\
     University of Wisconsin\\
     1150 University Ave\\
     Madison, Wisconsin 53706, USA\\
     \email{calin@physics.wisc.edu}\\
    }
\Abstract{
We study D-brane moduli spaces and tachyon condensation in 
B-type topological minimal models and their massive deformations. 
We show that any B-type brane is isomorphic with a direct sum of `minimal'
branes, and that its moduli space is stratified according to the type
of such decompositions. 
Using the Landau-Ginzburg formulation, we propose a closed formula for the
effective deformation potential, defined as the generating function of
tree-level open string amplitudes in the presence of D-branes. 
This provides a direct link to the
categorical description, and can be formulated in terms of holomorphic
matrix models. We also check that the critical locus of this potential
reproduces the D-branes' moduli space as expected from general
  considerations. Using these tools, we perform a detailed analysis of
  a few examples, for which we obtain a complete algebro-geometric description of 
 moduli spaces and strata. }
\date{}

\maketitle
\tableofcontents
\pagebreak

%%%%%%%%%%%%%%%%%%%%%%%%%

\section{Introduction}

%%%%%%%%%%%%%%%%%%%%%%%%%

A central problem in ${\cal N}=1$ string compactifications with D-branes is
the computation of superpotential terms arising from scattering of open strings.
While this is far from being solved in nontrivial set-ups (for
example D-branes on compact Calabi-Yau manifolds), one can gain important insights
by studying more basic building blocks such as $N=2$ minimal models. It is
by now well understood  that the superpotential is computed by the associated topological
string, and that it can be represented as the generating function of tree-level open
string amplitudes (see \cite{CIL4,HLL} for a detailed discussion of this
aspect). This allows one to approach the subject with the powerful methods
of topological string theory. Basic results in this respect were obtained in
\cite{HLL}, where it was shown that the generating function
of open string amplitudes (the so-called {\em effective potential} $\weff$) satisfies a
countable series of constraints which generalize the well-known associativity
equations of \cite{DVV}, and that it can be viewed as a (non-cubic)
string field action in an approximation in which the closed string variables
are treated as backgrounds. While the interpretation of $\weff$ as a
space-time superpotential only makes sense in critical string
theories, this quantity is well-defined for general
models, and the properties mentioned above are not restricted to the
critical case \cite{HLL}.

An important aspect of the effective potential is that it encodes
all obstructions to D-brane deformations, namely 
the true D-brane moduli space can be represented locally as
the critical set of $\weff$ in a space of linearized deformations
\cite{CIL4}. This connects obstruction theory to open string dynamics, 
and provides a tool for analyzing the local moduli
space. In many examples, the latter is an algebraic variety
and can be studied quite explicitly.

In the present paper, we study effective potentials and topological
D-brane moduli spaces from a global perspective, focusing on 
B-twisted minimal models and their massive
deformations\footnote{Some examples of boundary flows and 
tachyon condensation at the conformal
point of the parameter space were previously discussed in 
\cite{freden,cappelli,Horiflows}.}.
Upon using the Landau-Ginzburg realization of such theories
\cite{Vafa_LG,Labastida}, we 
propose a closed form for the effective potential, 
which expresses this quantity through the category-theoretic data of
\cite{Kontsevich,Orlov,Kap1,BHLS,Kap2,Kap3}. This 
generalizes results obtained by solving the consistency
conditions of \cite{HLL} in a series of examples, and should be viewed
as a conjecture still awaiting proof.  As a consistency check, 
we show that the (matrix) factorization locus of the Landau-Ginzburg superpotential coincides
locally with the critical locus of $\weff$. This concise proposal 
should be contrasted with the bulk WDVV (pre-)potential $\cal F$, for which
a closed expression  is not known. 
When combined with a general
characterization of moduli spaces, it allows for a complete 
description of the topological version of tachyon condensation in such
models.

Specifically, we shall consider minimal models of type $A_{k+1}$,
whose target space in the Landau-Ginzburg realization is the affine
line. In this case, the categorical description of $B$-type branes
(originally proposed by Kontsevich \cite{Kontsevich} and further developed in \cite{Orlov}) is
extremely simple. First, all projective modules over $\C[x]$ are
free, which allows us to represent branes as pairs $\cM=(M,D)$ where
$M=\Mp\oplus \Mm$ with $\Mp=\Mm=\C[x]^{\oplus r}$ and $D$ is a block
off-diagonal polynomial matrix of type $(2r)\times (2r)$.
Physically, $\Mp$ and $\Mm$ represent $r$ pairs of $D2$ branes and antibranes 
of the associated sigma model, which condense to configurations of
$D0$-branes when turning on the Landau-Ginzburg superpotential 
\cite{Kap1,BHLS,Kap2,Kap3,coupling,traces}. 
Moreover, $D$ models the boundary part of the BRST operator.  Second,
the ring of univariate polynomials is a PID (principal ideal domain), which
implies that each $\cM$ is isomorphic with a direct sum of
pairs $\cM_i=(M_i,D_i)$ for which $M_{+,i}=M_{-,i}=\C[x]$.  Such
`rank one' objects are the analogs of `rational D-branes' for massively
perturbed B-type minimal models, and we shall call them `minimal
branes'. 

Thus {\em every D-brane 
is isomorphic with a direct sum of minimal branes}.
This observation implies that a D-brane's moduli correspond to
varying the isomorphism between $\cM$ and its minimal brane decomposition
$\oplus_i \cM_i$,
and allows for a computable description of the boundary moduli
space. In fact, we will show that such moduli spaces are stratified
according to the minimal brane content. In particular, the minimal brane 
decomposition changes each time one crosses from a stratum to
another. Such transitions implement composite formation processes,
thus giving an explicit description of `topological tachyon
condensation'\footnote{In the topological version of tachyon condensation,
there are only topological (i.e. twisted) models for tachyons, and one deforms the theory
along flat directions rather than starting from the top of a Mexican
hat potential. Nevertheless, one encounters a truncation of the open
string spectrum, and composite formation along such deformations. For
critical topological string theories, this version of
tachyon condensation was discussed abstractly in \cite{CIL2,CIL3} and
analyzed in certain examples in \cite{CIL6,CIL7,CIL8,CIL9,CIL10}
with the tools of string field theory (see \cite{CIL5} for a review). }
in minimal models and their massive deformations.  This provides a
set of theories for which open string tachyon condensation and D-brane composite 
formation can be studied exactly.

The paper is organized as follows. In Section 2, we analyze 
general aspects of D-branes in minimal models and their massive deformations.
We first recall some basic
facts about D-branes in the Landau-Ginzburg formulation, focusing on the
category-theoretical description of \cite{Kontsevich,Orlov}. After
discussing the subcategory of minimal branes, we show that
every D-brane is isomorphic with a direct sum of such objects, 
a result which relies on the existence of a Smith normal form for univariate
polynomial matrices. We also determine the stratification of the moduli
space of a D-brane induced by minimal brane decompositions. Section 3
considers the physical description of D-brane deformations induced by
turning on the coupling to `odd' boundary  observables. After
explaining the type of deformations considered in this paper, we
discuss the parameterization of linearized deformations for
minimal branes. We also show how composites obtained
through topological tachyon condensation can be described as objects
of the original D-brane category, and determine the appropriate
parameterization of linearized deformations for such
composites. In Section 4, we discuss our proposal for the effective
potential $\weff$, starting with the case of minimal branes. 
After recalling the results obtained in \cite{HLL}, 
we cast them into a closed form which resembles a generalized residue
formula.  Subsection 4.2 extends this proposal to general B-type
branes and shows that the critical set of the effective potential
reproduces the deformation space. 
In Subsection 4.3, we show that our proposal for $\weff$ can be written as
the classical potential of a holomorphic matrix model as defined and
studied in \cite{holo}, and explain how the `constituent $D0$-branes'
of a minimal brane arise in this description. 
Section 5 discusses moduli spaces and tachyon condensation processes in
a series of examples. After recalling the physics-inspired parameterization 
introduced in Section 3, we give a detailed discussion of moduli
spaces for certain composites of minimal branes, which  we
describe completely as affine varieties. We also extract the associated strata, and
discuss appropriate reparameterizations for the effective potentials
along such strata.

%%%%%%%%%%%%%%%%%%%%%%%%%

\section{$D$-branes in topological minimal models}

%%%%%%%%%%%%%%%%%%%%%%%%%

In this section we discuss general aspects of D-branes in topological
minimal models and their massive deformations. After recalling the
description due to
\cite{Kontsevich,Orlov,Kap1,BHLS,Kap2,Kap3,coupling,traces}, 
we discuss the subcategory of minimal branes and
show that any brane is isomorphic with a sum of minimal branes. We
also show that the moduli space of a given D-brane is naturally
stratified according to its minimal brane content.

\subsection{Landau-Ginzburg description of B-type branes}
\label{LGB}

%%%%%%%%%%%%%%%%%%%%%%%%%

Let us recall the construction of B-type branes in
$A_{k+1}$ topological minimal models. The bulk sector is described by a
twisted $N=2$ minimal model at $SU(2)$ level $k$,
together with its massive deformations. 
These theories have a convenient Landau-Ginzburg
realization \cite{Vafa_LG,Labastida,DVV}, in which 
the bulk sector is characterized by the superpotential:
\beql{wlg}
\wlg(\vp,t)\ =\ \frac{\vp^{k+2}}{k+2}-\sum_{i=0}^{k} g_{k+2-i}(t)\, \vp^i\ .
\eeql
We will often write this polynomial in terms of its distinct roots $\tilde x_i(t)$:
\beq
\label{Wfact}
W(x;t)=\frac{1}{k+2}\prod_{i=1}^{s}{(x-{\tilde x}_i(t))^{k_i}}~~,
\eeq
where $1\leq s \leq k+2$ and $k_i\geq 1$ with $\sum_{i=1}^s{k_i}=k+2$. 
The polynomials $g_{k+2-i}(t)$ in \rf{wlg} depend on flat coordinates $t_i$
($i=2\ldots k+2$) as described in \cite{DVV}. These 
coordinates measure the strength of couplings in the perturbed bulk action,  
namely:
\beq
\label{Sbulk_var}
\delta S=\sum_{i=0}^k t_{k+2-i}\int d^2z \{G_{-1},[{\bar
G}_{-1},\Phi_i]\}~~,
\eeq
where $G_{-1}$ and ${\bar G}_{-1}$ are modes of the left and right moving
twisted supercurrents. The case $t=0$ (i.e., $W=\frac{x^{k+2}}{k+2}$) 
corresponds to the twisted $N=2$ superconformal minimal model,
while general values of $t_i$ describe its massive topological deformations. 
The observables $\Phi_i$ form a linear basis 
of the space of chiral primary fields, which can be written as the Jacobi algebra
of the Landau-Ginzburg superpotential: 
\beq
{\cR}:=\BC[\vp]/\langle\del_\vp\wlg\lab{k+2}(\vp)\rangle~~.
\eeq
Thus one can choose the basis:
\beql{chiralring}
\Phi_i=\vp^i~~,~~\qquad i=0\ldots k~~,
\eeql
which corresponds to the parameterization used in \rf{wlg}.

A general analysis of the boundary sector of B-twisted Landau-Ginzburg
models was performed in \cite{Kap1,BHLS,Kap2,Kap3,coupling,traces},
upon using a suggestion of \cite{Kontsevich}.\footnote{See 
\cite{Warner,Govindarajan:1999js,Govindarajan:2000my,HIV,Horilin,Govindarajan} for 
previous work on the general subject of D-branes in Landau-Ginzburg
models, and \cite{mirrbook} for a review.}
The situation is simplest for single, `minimal' branes
(whose precise definition will be given below).  Such branes
correspond to all polynomial factorizations of the
bulk superpotential:
\beql{JEW}
W(x)\ =\ J(x)\,E(x)~~,
\eeql
where $J(x)$ plays the role of a boundary superpotential
\cite{BHLS}. As discussed in more detail in 
Section 4.3, the 
zeroes of $J(x)$ can be viewed as locations of $D0$ branes in the target space $\C$.
Moreover, $E$ appears in the generalized chirality condition for fermionic
boundary superfields \cite{Govindarajan,Hellerman:2001bu}.
The generalization to arbitrary D-branes is obtained by
promoting  $J$ and $E$ to square polynomial matrices (square matrices
whose entries belong to the polynomial ring $\C[x]$). These describe 
\cite{Horilin,Kap1,BHLS,Kap2,Kap3,coupling,traces}
the tachyon profiles for a set of pairs of D2-branes and antibranes 
of the associated {\em sigma} model, which is recovered by
turning off $W$.  

In more abstract
language, a general topological B-type brane is described by 
the pair of data $\cM=(M,D)$, where $M$ is
a free $\C[x]$-supermodule and $D$ an odd module endomorphism
subject to the condition\footnote{A
  slightly more general construction is allowed \cite{coupling,traces},
  but we shall not consider it here. This amounts to adding a
  constant operator to the right hand side of (\ref{Dsquare}).}:
\beq
\label{Dsquare}
D^2=W\bfone~~.
\eeq
The endomorphism $D$ plays the role of a boundary BRST operator
\cite{Kap2,traces}, and can be viewed as a square polynomial 
matrix upon choosing an arbitrary basis of $M$. 
Considering the homogeneous decomposition $M=\Mp \oplus
\Mm$, which corresponds to a decomposition into sigma model $D2$ branes and
antibranes, it is clear that (\ref{Dsquare}) has no solutions unless
$\rk \Mp=\rk \Mm$, so we will always assume that this condition
holds and denote the common rank by  $r$. Using this decomposition,
we write:
\beq
\label{Dmatrix}
D=\left[\begin{array}{cc}0&E\\J&0\end{array}\right]~~,
\eeq
with $J\in \Hom_{\C[x]}(\Mp,\Mm)$ and $E\in \Hom_{\C[x]}(\Mm,\Mp)$. 
Expression (\ref{Dmatrix})  brings (\ref{Dsquare}) to the form: 
\beq
\label{factcond}
J\,E\ =\ E\,J\ =\ W\,\bfone~~,
\eeq
which describes factorizations of the polynomial $W$ into square polynomial
matrices of type $r\times r$. 

In this construction, 
the space of boundary topological observables 
is modeled by ${\cal D}$-cohomology,
where ${\cal D}$ is the nilpotent operator on $\End_{\C[x]}(M)$ defined by taking the
supercommutator of an endomorphism with $D$: 
\beq
{\cal D}=[D,\,\cdot\,]~~.
\eeq
With the product induced by composition of endomorphisms, the space
$H_{\cal D}(\End_{\C[x]}(M))$ becomes a superalgebra over $\C$. Then the
space of boundary observables can be identified with $H_{\cM}:=G\otimes_\C
H_{\cal D}(\End_{\C[x]}(M))$, where $G$ is a Grassmann algebra. 
An explicit analysis of ${\cal D}$-cohomology for minimal branes in minimal models and
their massive deformations was performed in \cite{BHLS}, 
and some of the results will be recalled below.

For general $D$-brane configurations, one also has boundary-condition
changing observables which correspond to excitations of strings stretching
between pairs of D-branes. Viewing these as morphisms between branes leads
naturally to the category-theoretic picture expected from the work of
\cite{CIL1,Moore_Segal,Moore,Douglas,CIL2,CIL3,Aspinwall}. 
In Landau-Ginzburg models, the precise realization of this description
was proposed by Kontsevich \cite{Kontsevich} and 
developed by Orlov \cite{Orlov}.  As expected from
general considerations, the collection of all B-type
branes forms an (enhanced) triangulated category $\DB_W$,
which in the present case is only $\Z_2$-graded. It arises
as the cohomology category of a differential graded category $\DG_W$, the
so called 'Kontsevich category of pairs', which encodes off-shell tree-level
open string data. By contrast, the cohomology category $\DB_W$ encodes the on-shell information. 

The objects of $\DG_W$ are branes $\cM=(M,D_M)$ as described above, viewed  
as pairs of $\C[x]$-module morphisms: 
\beql{Pdef}
\cM\ =\ \Bigl(
\xymatrix{
\Mm\ \ar@<0.6ex>[r]^{E} &\ \Mp \ar@<0.6ex>[l]^{J}}
\Bigl)~~
\eeql
subject to condition (\ref{factcond}). Morphism spaces are defined by:
\beq
\Hom_{\DG_W}(\cM,\cN)=\bigoplus_{\alpha,\beta=+,-}{
\Hom(M_\alpha,N_\beta)}~~, 
\eeq
with the obvious $\Z_2$ grading and the differential:
\beq
{\cal D}_{M,N}(f)=D_N\circ f-(-1)^{|f|}f\circ D_M~~,
\eeq
where $|\cdot|$ denotes the degree of homogeneous morphisms.
Boundary preserving observables in the sector $\cM$ correspond to elements of
$H_{{\cal D}_{M,M}}(\Hom(\cM,\cM))$, 
while boundary condition changing states are elements of  
$H_{{\cal D}_{M,N}}(\Hom(\cM,\cN))$ for $\cM\neq \cN$. 

The `on-shell' data is recovered as the total 
cohomology category $\DB_W:=H(\DG_W)$, whose objects coincide with those of
$\DG_W$ and whose morphism spaces are obtained by passing to cohomology in
each $\Hom_{\DG_W}(\cM,\cN)$. This category is triangulated\footnote{More
  precisely, the subcategory $H^0(\DG_W)$ of $\DB_W$ is triangulated in the standard sense and
  $H^1(\DG_W)=H^0(\DG_W)[1]$ where $[1]$ is the shift functor discussed below.}
due to mathematical results of
\cite{BK}. The construction is very similar in spirit to that of 
\cite{CIL2,CIL3} (see \cite{CIL5} for a review and
\cite{CIL6,CIL7,CIL8,CIL9,CIL10} for applications to
twisted sigma models). A different point of view, which does not rely
directly on the results of \cite{BK}, was discussed in \cite{Douglas,Aspinwall}. 

It is also useful to introduce an intermediate category $\P_W$ such
that $Ob \P_W=Ob \DG_W=Ob \DB_W$ and
$\Hom_{\P_W}(\cM,\cN)=Z(\Hom_{\DG_W}(\cM,\cN)):=
\{f\in \Hom_{\DG_W}(\cM,\cN)|{\cal D}_{M,N}(f)=0\}$, the
space of cocycles in the complex $ \Hom_{\DG_W}(\cM,\cN)$. The latter
has the decomposition $\Hom_{\P_W}(\cM,\cN)=\Hom^0_{\P_W}(\cM,\cN)\oplus
\Hom^1_{\P_W}(\cM,\cN)$, where $\Hom^\alpha_{\P_W}(\cM,\cN)=\{f\in
\Hom_{\P_W}(\cM,\cN)| f={\rm ~homogeneous~and~}|f|=\alpha\in \Z_2
\}$. Restricting morphism spaces to even components gives a
subcategory $\P_W^0$ for which 
$\Hom_{\P_W^0}(\cM,\cN)=\Hom_{\P_W}^0(\cM,\cN)$. Notice that
two objects $\cM,\cN$ are isomorphic in $\P_W^0$ if and only if there
exists an even module {\em isomorphism} $U\in \Hom^0_{\C[x]}(M,N)$ such
that $D_N\circ U=U\circ D_M$; this is the natural notion of
isomorphism between the branes $\cM$ and $\cN$.  

Antibranes are described by
acting with the shift functor $[1]$, whose main effect is to flip $J$ and $E$:
\beql{shift}
\cM[1]\ =\ \Bigl(
\xymatrix{
\Mp \ \
\ar@<0.6ex>[r]^{-J} &
\ \ \Mm 
\ar@<0.6ex>[l]^{-E}
}\Bigl)\ ~~
\eeql
while acting in standard fashion on morphisms:
\beq
f=(f_0, f_1)\rightarrow f[1]=(f_1,f_0)~~. 
\eeq
Since the categories are only $\Z_2$-graded, we have $[1]^2={\rm id}$, where ${\rm id}$ is
the identity functor.

%%%%%%%%%%%%%%%%%%%%%%%%%%%%%%%%%
\subsection{The subcategory of minimal branes}
%%%%%%%%%%%%%%%%%%%%%%%%%%%%%%%%%

The simplest class of objects in $\DG_W$ is obtained by choosing $\rk \Mp=\rk
\Mm=1$, i.e. $\Mp=\Mm=\C[x]$. Then $J$ and $E$ are polynomials satisfying
(\ref{factcond}), and they are easily described by using the factorization
(\ref{Wfact}) of $W$. Namely 
\beq
\label{minimal_gen}
J(x)=C\prod_{i=1}^{s}{(x-{\tilde x}_i(t))^{m_i}}~~,
~~E(x)=\frac1{C(k+2)}\prod_{i=1}^{s}{(x-{\tilde x}_i(t))^{k_i-m_i}}~~,
\eeq
where $C$ is a non-vanishing complex constant and the integers $m_i$ satisfy
$0\leq m_i\leq k_i$. We let: 
\beq
\ell+1:=\deg J=\sum_{i=1}^s{m_i}~~.
\eeq
When $C=1$ (i.e. when $J$ is a monic polynomial), 
each pair (\ref{minimal_gen})  is characterized 
by the integral vector ${\bf m}:=(m_1\dots m_s)$ and 
defines objects $\cM_{\bf m}$ of $\DG_W$ which we shall call 
{\em minimal branes}. The collection of such objects defines
a full subcategory of $\DG_W$, the {\em minimal subcategory}. 
Notice that we keep the constant $C$ explicitly, 
since we defined the objects
of $\DG_W$ to be pairs $(M,D)$ and not classes of such
pairs up to a rescaling\footnote{Since we are in a quantum theory, 
rescaling by nonzero constants is not physically relevant (only
the Hilbert space ray of any state is physically meaningful). 
One can
implement this by passing to the category whose objects are
equivalence classes of pairs $(M,D)$ under such rescalings. Since 
this would clutter the notation, we prefer to use
the original objects $(M,D)$, with the understanding that such
rescalings do not affect the underlying physics.}. 
Moreover, notice that minimal branes are defined to have $C=1$.  
The total number
of minimal branes equals $\prod_{i=1}^s{(k_i+1)}$.

The boundary preserving and boundary condition changing spectra of minimal branes
were computed in \cite{BHLS}. The boundary preserving spectrum,
described by 
$H_{\cal D}(\End_{\C[x]}(M))$, forms a super-commutative algebra $\BC[\vp,\omega]/{\mathcal I}$ 
with even and odd generators $\vp$ and $\omega$. The ideal of 
relations can be described as follows.
Let $G$ denote the greatest common divisor of $J$ and $E$, whose degree we
denote by $l+1$. We have:  
\beql{JpG}
J=p\,G ~{\rm and}~E=q\,G
\eeql
for some coprime polynomials $p$ and $q$ in the variable $x$. 
Then the odd generator is\footnote{The convention differs slightly from
  that of \cite{BHLS}, and is fixed by agreement with \rf{Jdef} and
  (\ref{Dlin}).}: 
\beq
\omega=\left[\begin{array}{cc}0&q\\-p&0\end{array}\right]~~,
\eeq
and the ideal ${\mathcal I}$ is generated by the elements:
\beql{boundideal}
G(\vp)~{\rm and}~~\omega^2+p(\vp) q(\vp)~~.
\eeql
To these generators we can associate rational degrees, which play the
role of $U(1)$ charges at the superconformal point $t=0$:
\beql{bpcharges}
q(\vp)\ =\ 1,\qquad q(\omega)\ =\ k/2-l~~.
\eeql
Corresponding to the split
$H_{\cal D}(\End_{\C[x]}(M))=H^0_{\cal D}(\End_{\C[x]}(M))\oplus H^1_{\cal D}(\End_{\C[x]}(M))$,
one finds the homogeneous basis:
\beql{boundring}
 \big\{\phi_\al,\psi_\al\big\}=
\big\{\vp^\al,\omega \vp^\al\big\}~~,~~
\al=0\ldots l~~.
\eeql

At the conformal point $t_i=0$, we have $W=\frac{x^{k+2}}{k+2}$ and we
can take $J=x^{\ell+1}$ and $E=\frac{x^{k+1-\ell}}{k+2}$.  Then $G=J$,
$p=1$ and $q=\frac{x^{k-2\ell}}{k+2}$.  This corresponds to $s=1$,
$k_1=k+2$ and $l=\ell$, with $\ell \in\{ -1\dots k+1\}$.  The number
of minimal branes equals $k+3$, and they will be denoted by
$\cM_\ell\in Ob\DG_W$. As shown in \cite{BHLS}, the branes $\cM_\ell$
with $\ell=0\ldots [k/2]$ correspond to the well-known `rational'
$B$-type boundary states of the $A_{k+1}$ minimal model. The object
$\cM_{-1}$ is the trivial brane, which can be physically identified
with the closed string vacuum\footnote{More precisely, we have
$\Hom_{\DB_W}(\cM_{-1}, \cM)=\Hom_{\DB_W}(\cM, \cM_{-1})=0$ for any
D-brane $M$. Thus $\cM_{-1}$ has trivial boundary preserving spectrum
and trivial boundary changing spectrum with any other brane.}.  The
objects $\cM_\ell$ with $\ell=[k/2]\ldots k+1$ are the `rational'
antibranes, due to the relation $\cM_\ell\approx \cM_{k-\ell}[1]$ (for
even $k$, the brane $\cM_{k/2}$ is isomorphic with its antibrane).
The choice $C=1$ corresponds to a particular normalization of the
`rational' boundary states.

The boundary condition changing spectrum associated with open strings stretching
between two minimal branes, is more complicated and we refer the
reader to \cite{BHLS} for its general description. 
The simplest form is found at the conformal point $t_i=0$,
where $\Hom_{\DB_W}(\cM_{\ell_1},\cM_{\ell_2})$ admits the following
basis:
\beql{bcring}
\big\{\phi_\gamma^{\ell_1,\ell_2},\psi_\gamma^{\ell_1,\ell_2}\big\}=
\big\{\beta\lab{\ell_1,\ell_2}\vp^\gamma,\omega\lab{\ell_1,\ell_2} \vp^\gamma\big\}~~,
~~\gamma=0\ldots\ell_{12}:= {\rm min}(\ell_1,\ell_2)~~.
\eeql
Its even and odd generators:
\beq
\beta\lab{\ell_1,\ell_2}=
\frac{1}{x^{\ell_{12}}}\left[\begin{array}{cc}x^{\ell_2}&0\\0&x^{\ell_1}\end{array}\right]~~,~~
\label{omega_bc}
\omega\lab{\ell_1,\ell_2}=
\left[\begin{array}{cc}0&x^{k-\ell_1-\ell_2}\\1&0\end{array}\right]~~,
\eeq
have $U(1)$ charges given by:
\beql{bccharges}
q(\beta\lab{\ell_1,\ell_2})\ =\ \frac12|\ell_1-\ell_2|~~,~~
q(\omega\lab{\ell_1,\ell_2})\ =\ \frac12(k-\ell_1-\ell_2)\ .
\eeql

%%%%%%%%%%%%%%%%%%%%%%%%%%%%%%%%%%%%%%%%%%%%
\subsection{Minimal brane decompositions}
\label{minimal_dec}
%%%%%%%%%%%%%%%%%%%%%%%%%%%%%%%%%%%%%%%%%%%%

Returning to general $D$-branes, 
let us fix a $\C[x]$-supermodule $M=\Mp\oplus \Mm $ with ranks given by
$\rk \Mp=\rk\Mm:=r$. 
Then the factorization equation (\ref{Dsquare})
is invariant under transformations of the form: 
\beq
\label{simaction}
D\rightarrow UDU^{-1}
\eeq
where $U\in \Aut^0_{\C[x]}(M)$ is an even module automorphism of $M$. Writing 
$D=\left[\begin{array}{cc}0&E\\J&0\end{array}\right]$ and:
\beq
U=S\oplus T=\left[\begin{array}{cc}S&0\\0&T\end{array}\right]~~,
\eeq
with $S\in \Aut_{\C[x]}(\Mp)$ and $T\in \Aut_{\C[x]}(\Mm)$, 
we find that (\ref{simaction}) amounts to the double similarity transformation:
\begin{eqnarray}
E&\rightarrow& SET^{-1}\nn\\
J&\rightarrow& TJS^{-1}~~.
\end{eqnarray}
The operators $S,T$ can be viewed as unimodular polynomial matrices, 
i.e. polynomial matrices which are invertible over $\C[x]$. 
This amounts to the requirement that their
determinants are units in the polynomial ring, i.e. $\det S$ and $\det T$ are nonzero
complex constants. 

Since $\C[x]$ is a PID, the matrix $J$ can always be brought to
Smith form by a double similarity transformation. Namely, we have: 
\beq
\label{Jsim}
J=TJ_cS^{-1}~~
\eeq
for some invertible $T$ and $S$, with: 
\beq
\label{Jsmith}
J_c={\rm diag}(p_1\ldots p_r)~~,
\eeq
where $p_j\in \C[x]$ are monic univariate polynomials satisfying the division relations: 
\beq
p_1|p_2|\ldots |p_r~~.
\eeq
These are given explicitly by:
\beq
\label{ps}
p_i:=\frac{G_i}{G_{i-1}}~~\forall i=1\dots r~~,
\eeq
where $G_i$ is the monic greatest common divisor of all $i\times i$ minors of
$J$, with the convention $G_0:=1$. 
It is clear that $\frac{W}{J}$ is a polynomial matrix if and only if $p_r|W$. 

Hence a solution $D$ of (\ref{Dsquare}) can always be brought to the form: 
\beq
D_c=\left[\begin{array}{cc}0&E_c\\J_c&0\end{array}\right]~~,
\eeq
where $J_c=\diag(p_1\dots p_r)$ and $E_c=\diag(q_1\dots q_r)$, with: 
\beq
q_i:=\frac{W}{p_i}\in \C[x]~~.
\eeq
Thus $D=UD_cU^{-1}$ for some unimodular matrix $U$, where 
$D_c=D_1\oplus \ldots \oplus D_r$, with: 
\beq
D_i:=\left[\begin{array}{cc}0&q_i\\ p_i&0\end{array}\right]~~. 
\eeq 
This shows that any D-brane $\cM=(M,D)$ is equivalent with the direct sum
of minimal branes $\cM_{1}\oplus \ldots \oplus \cM_{r}$, where 
$\cM_i:= 
(M_i,D_i)=(\C[x]{\scriptsize \begin{array}{c} q_i \\ \rightleftarrows
    \\  p_i\end{array}}\C[x])$,
with $M_{+,i}=M_{-,i}:=\C[x]$.
The equivalence is implemented by the module isomorphism $U^{-1}$, 
which has the property $U^{-1}D=(\oplus_{i}{D_{i}})U^{-1}$ (showing
that $U^{-1}$ is an isomorphism in the category $\P_W^0$). 
The direct  sum $\oplus_{i}{\cM_i}$ will be called  the {\em minimal
  brane decomposition} of $\cM$. We obtain the following\footnote{
This, of course, does not mean that each $(M,D)$ is a direct sum, since isomorphic objects 
in $\P_W^0$ cannot be identified in general. By definition the objects
of $\DG_W$, $\P_W$ and $\DB_W$
are pairs $(M,D)$ satisfying $D^2=W\bfone$, and {\em not} isomorphism classes
of such pairs.}:

\paragraph{\bf Proposition} {\em Any D-brane $\cM$ is isomorphic in
  $\P_W^0$ with a direct sum of minimal branes. In particular, the
finite subcategory of minimal branes generates $DB_W$ as an additive
category, up to isomorphisms.} 

While the target space description of the isomorphism is elementary,
we mention that the transformation $D\rightarrow UDU^{-1}$ amounts to a nontrivial change
of variables in the world-sheet action. This is because the boundary coupling
of \cite{coupling} contains terms dependent on $D(\phi(\tau))$, where
$\phi(\tau)$ is the restriction of the scalar field to the boundary of the
world-sheet. The microscopic transformation  
$D(\phi(\tau))\rightarrow U(\phi(\tau))D(\phi(\tau)) U(\phi(\tau))^{-1}$
induces a nonlinear change of function in the world-sheet action. Despite
this fact, the proposition shows that any 
topological B-brane is isomorphic with a direct sum of topological
minimal branes. This is a consequence of the existence of the Smith normal form,
which itself follows from the fact that our model's target space is $\C$.

Our category-theoretic discussion can be summarized by the following
sequence of operations:
\beql{sequence}
\begin{array}{ccccc}
\DG_W &\stackrel{{\rm restrict~to~cocycles}}{\longrightarrow}&
  \P_W  &\stackrel{{\rm divide~by~homotopies}}{\longrightarrow}&  \DB_W\\ 
~&~& \downarrow&~&\uparrow\\
~&~&{\rm M}_W&\stackrel{{\rm
  divide~by~homotopies}}{\longrightarrow}& {\rm DM}_W
\end{array}
\eeql
where the right vertical arrow denotes inclusion and the left vertical
arrow  means that  we identify objects which differ by isomorphisms in
    $P_W^0$ (this induces corresponding identifications of the
    morphism spaces).

In physical terms the horizontal arrows implement passage to the 
cohomology of the boundary BRST-operator, in order to obtain
the physical spectrum of the associated $B$-type brane. The category
$M_W$ is the direct sum completion of the category of minimal branes,
while $DM_W$ is its `derived category'. The latter coincides with $DB_W$ up
to isomorphisms.

%%%%%%%%%%%%%%%%%%%%%%%%%%%%%%%%%%%%%%%%%%%
\subsection{The minimal brane stratification of the factorization locus}
%%%%%%%%%%%%%%%%%%%%%%%%%%%%%%%%%%%%%%%%%%%

For fixed closed string moduli $t_i$ and rank $r$, the factorization
condition (\ref{Dsquare}) will generally admit families
of solutions.  Let us denote the space of solutions by ${\cal
S}_t$ and discuss some of its properties.  Later on we
will briefly consider the union ${\cal Z}=\sqcup_{t}{{\cal
S}_t}$, which is obtained by allowing $t$ to vary.  

Let ${\cal M}$ denote the quotient of ${\cal
S}_t$ by the action of the similarity transformation (\ref{simaction}).
  We have the orbit decomposition:
\beq
{\cal S}_t=\sqcup_{(p_1\ldots p_r) \in {\cal M}}{{\cal O}_{\tiny (p_1\ldots p_r)}}~~.
\eeq
It is clear from Section \ref{minimal_dec} that  ${\cal M}$ consists of monic
tuples $(p_1\ldots p_r)\in \C[x]^r$, subject to the divisibility constraints:
\beq
\label{divcond}
p_1|p_2|\ldots |p_r|W~~.
\eeq
Thus:
\beq
\label{calMp}
{\cal M}=\Big{\{}(p_1\ldots p_r)\in \C[x]^r~\Big{|}~p_1|p_2|\ldots |p_r|W~{\rm and}
~{\rm lcoeff}(p_j)=1~{\rm for~all}~j=1\ldots r~\Big{\}}~~.
\eeq
To make this more explicit, recall the factorization (\ref{Wfact}) of the bulk
Landau-Ginzburg superpotential.
Writing
$p_j(x)=\prod_{i=1}^{s}{(x-{\tilde x}_i(t))^{m_{i,j}}}$ leads to the following description of
the set of orbits:
\beq
\label{calMm}
{\cal M}\equiv
\{{\hat m}=(m_{ij})\in Mat(s,r;\Z)~|~0\leq m_{i1}\leq m_{i2}\leq \ldots \leq m_{ir}\leq
k_i~\forall i=1\dots s~\}~~.
\eeq
In particular, we can index orbits by the integral matrix ${\hat m}$. The orbit
${\cal O}_{\hat m}$ has the form:
\begin{eqnarray}
{\cal O}_{\hat m}=\Big{\{} D=\left[\begin{array}{cc}0&SE_cT^{-1}\\
    TJ_cS^{-1}&0\end{array}\right]
~&\Big{|}&~J_c={\rm diag}(\prod_{i=1}^{s}{(x-{\tilde x}_i(t))^{m_{i1}}} , \ldots , 
\prod_{i=1}^{s}{(x-{\tilde x}_i(t))^{m_{ir}}})~~,\nn\\
& &~E_c={\rm diag}(\prod_{i=1}^{s}{(x-{\tilde x}_i(t))^{k_i-m_{i1}}} , \ldots ,  
\prod_{i=1}^{s}{(x-{\tilde x}_i(t))^{k_i-m_{ir}}})~~,\nn\\
& &~S,T\in GL(\C[x],r)
\Big{\}}~~.\nn
\end{eqnarray}
It is clear that 
$\cM_j:= (\C[x]{\scriptsize \begin{array}{c} q_j \\ \rightleftarrows \\
    p_j\end{array}}\C[x])$ is the minimal brane 
    $\cM_{{\bf m}^{(j)}}$, where 
${\bf m}^{(j)}=(m_{1j}\dots m_{sj})$ is the $j$'th column of the integral matrix
${\hat m}$. Thus branes belonging to the stratum ${\cal O}_{\hat m}$ are isomorphic
with the direct sum $\cM_{{\bf m}^{(1)}}\oplus \ldots \oplus \cM_{{\bf m}^{(r)}}$. 
This gives the following:

\paragraph{\bf Proposition}{\em The strata ${\cal O}_{\hat m}$ of the solution space
  ${\cal   S}_t$ are characterized by different minimal brane decompositions.} 

Hence the minimal content of a brane `jumps' along its deformation space each
time one crosses from one stratum to another.

%%%%%%%%%%%%%%%%%%%%%%%%%%%%%%%%%%%%%%%%%%%
\section{Deformations of $D$-branes by boundary operators}
%%%%%%%%%%%%%%%%%%%%%%%%%%%%%%%%%%%%%%%%%%%

%%%%%%%%%%%%%%%%%%%%%
\subsection{A first look at linearized deformations and obstructions}
\label{lindef}
%%%%%%%%%%%%%%%%%%%%

Let us consider a minimal brane $(M,D)\in Ob \DG_W$ and focus for
simplicity on its boundary deformations. 
A basis of the even and odd components of
$H_{\cal D}(\End_{\C[x]}(M))$ can be chosen as in \rf{boundring}.
This defines (Grassmann-valued) linear coordinates $(\xi_\alpha,
\eta_\alpha)$ on $H_M=G\otimes_{\C}H_{\cal D}(\End_{\C[x]}(M)) $, 
allowing us to write an arbitrary boundary observable ${\cal B}$ in the form:
\beql{bobs}
{\cal B}=\sum_{\alpha}{\xi_\alpha \phi_\alpha}+\sum_\alpha{\eta_\alpha\psi_\alpha}~~.
\eeql
It is clear that $\xi_\alpha$ have the same $\Z_2$-degree as
${\cal B}$, while $\eta_\alpha$ have opposite degree. 
The observables ${\cal B}$ can be used to deform the
world-sheet action. Because the latter is constructed from the
tachyon condensate \cite{Kap1,BHLS,Kap2,Kap3,coupling,traces}, the simplest
deformations correspond to:
\beq
\label{Dlin}
D\rightarrow D':=D+\sum_{\alpha}{\xi_\alpha \phi_\alpha}+\sum_\alpha{\eta_\alpha\psi_\alpha}~~,
\eeq
with even $\eta_\alpha$ and odd $\xi_\alpha$ (since the observable $1_G\otimes D\equiv D$ is odd). 
In this expression, $\psi_\alpha$ and $\phi_\alpha$  denote representative cocycles
of the associated ${\cal D}$-cohomology classes.
Noticing that $D$ enters linearly\footnote{More precisely, the term in the exponent of the path-ordered
exponential of \cite{coupling} depends linearly on $D$. 
As explained in \cite{traces}, it is possible to add a supplementary
  term $K$ to the boundary action,  which depends quadratically on $D$ (such a
  term was also considered in \cite{Kap1,BHLS,Kap3} in special cases). 
While this term does not affect BRST invariance of the world-sheet
partition function, it is required if one wishes to have invariance
under both generators of the untwisted $N=2$ algebra. Since 
$\weff$ is a generating function for
tree-level {\em topological} string amplitudes, we can restrict to
the B-twisted model. In this case, the quadratic term $K$ does
not play any role, and can be ignored 
(as explained in \cite{traces}, this term does not contribute to 
localized correlators). The situation is similar to that encountered in
\cite{Witten_mirror} for B-type sigma models.}
in the boundary action $S_\del$ of \cite{coupling}, we find that this variation amounts to
the following deformation: 
\beq
\label{Sbdef}
\delta S_\del =
\sum_\alpha {\xi_\alpha \int_{\partial \Sigma}{d\tau~[G^b_{-1},\phi_\alpha]}}+
\sum_\alpha{\eta_\alpha\int_{\partial \Sigma}{d\tau~\{G^b_{-1},\psi_\alpha\}}}~~,
\eeq
where $G^b_{-1}$ are modes of the twisted boundary
supercurrent and $\partial \Sigma$ is the boundary of the world-sheet.
Notice that the insertion of $G_{-1}^b$ flips the total
$\BZ_2$ degree, so odd basis elements are associated with 
the even deformation parameters $\eta_\alpha$, while even basis elements are
associated with the odd parameters $\xi_\alpha$. 
As pointed out in \cite{HLL}, the odd
deformation parameters $\xi_\alpha$ drop out of the effective
potential in the boundary preserving sector.\footnote{Keeping them would require working with non-commuting
supercoordinates in order to prevent their cancellation in $\weff$
due to graded symmetrization \cite{HLL}.} Accordingly, we will focus on deformations
induced by turning on $\eta_\alpha$, which also have a simpler physical
interpretation.

Since $\eta_\alpha$ enter linearly in (\ref{Sbdef}), 
they are analogous to the bulk flat coordinates $t_i$ appearing in
(\ref{Sbulk_var}). Thus one expects $\eta_\alpha$ to play a role similar to that 
known for $t_i$ from the theory of Frobenius manifolds. While we shall not attempt to
do this here, we mention that one can cast the
results of \cite{HLL} in terms of a certain non-commutative version of
Frobenius manifolds, a fact which should be used to give an intrinsic
characterization of such coordinates. 
For the purpose of the present paper, we shall view $\eta_\alpha$
simply as linear coordinates on the vector space $H^1_{\cal D}(\End_{\C[x]}(M))[1]$.

Deformations of type (\ref{Dlin}) are generally obstructed,
since $D'$ need not satisfy the condition $D'^2=W\bfone$. Writing $D'=D+\delta
D$, the integrability equation takes the Maurer-Cartan form: 
\beq
\label{MC}
(\delta D)^2+[D,\delta D]=0~~.
\eeq
This can be studied via methods of algebraic homotopy theory as in
\cite{CIL4}. Some basic aspects of this were recently discussed in
\cite{hori_inf,Diac_inf} in the context of topological Landau-Ginzburg orbifolds.

As explained in \cite{CIL4} and \cite{HLL} in a more general context, the effect of obstructions
is encoded by the effective potential $\weff$, which in our case is defined on 
$H^1_{\cal D}(\End_{\C[x]}(M))$ and coincides with the 
generating functional of tree-level open string amplitudes in the boundary
sector described by the brane $\cM$. Then the even part ${\cal S}_t^{\cM}$ of the deformation space
of $\cM$  (i.e. the space of those solutions to 
(\ref{factcond}) or (\ref{MC}) which are continuously connected to the
reference solution $D$) can be {\em locally} described as the
critical set of $\weff$: 
\beq
{\cal S}_t^{\cM}\approx_{locally}\{\eta| \partial_{\eta_\alpha}\weff= 0~{\rm for~all~}\alpha \}~~.
\eeq
This gives a local realization of ${\cal S}_t^{\cM}$ as an affine variety inside $H^1_{\cal
  D}(\End_{\C[x]}(M))$. Of course, the deformation space ${\cal S}_t^{\cM}$ of $\cM$ is a subset of
the full solution space ${\cal S}_t$.

%%%%%%%%%%%%%%%%%%%%%%%%%%%%%%%%%%%%%%%%%%%
\subsection{Deformations of minimal branes}
%%%%%%%%%%%%%%%%%%%%%%%%%%%%%%%%%%%%%%%%%%%

It is clear from equations (\ref{minimal_gen}) that a minimal brane
$\cM_{\bf m}$ has a single boundary deformation parameter, namely the
non-vanishing constant $C$. However, such branes have $l+1$ linearized
deformations, associated with the basis \rf{boundring} of boundary
observables. Thus $l$ linearized
deformations must be obstructed due to the effective potential. 

Let us fix some closed string
moduli $t_i$, for which $\cM_{\bf m}$ is specified by the solution
$(J_0(x;t),E_0(x;t))$ of the factorization condition \rf{JEW}. 
In the basis \rf{boundring},
the linear parameterization of tachyon deformations takes the form:
\beq
\label{Jdef0}
J(\vp;t,\eta)\ =\ J_0(\vp;t)-\sum_{\al=0}^{l}\eta_\alpha\,\vp^\al~~,
~~\ell=0\dots [k/2]~~,
\eeq
with $l+1=\deg (\gcd(E,F)) \leq \ell+1=\deg J_0(x)$.
As mentioned in the previous subsection, we turn on only 
the even deformation parameters $\eta_\alpha$. 
Notice that (\ref{Jdef0}) ignores the modulus provided by the constant 
$C$ in (\ref{minimal_gen}), which can be taken trivially into account. Thus -- once
the effect of obstructions is implemented -- we should find a reduced moduli space
which consists of a single point. 

Writing $J_0(x;t)=x^{\ell+1}+\sum_{\alpha=0}^\ell{a_\alpha(t)
  x^\alpha}$, we find:
\beql{Jdef}
J\lab{\ell+1}(\vp;u)\ =\ \vp^{\ell+1}-\sum_{\al=0}^{\ell}
u_{\ell+1-\al}\,\vp^\al~~,
~~\ell=0\dots [k/2]~~,
\eeql
where we introduced the shifted parameters: 
\begin{eqnarray}
u_{\ell+1-\al}&:=&\eta_\alpha-a_\alpha(t)~~{\rm for}~~\alpha=0\dots l\nn\\
u_{\ell+1-\al}&:=&-a_\alpha(t)~~{\rm for}~~\alpha=l+1\dots \ell~~,\nn
\end{eqnarray}
Concentrating on linearized deformations
of $\cM_{\bf m}$ means that one allows only $u_{\ell+1-l}\ldots u_{\ell+1}$ to
vary. However, it is convenient to permit variations of all $u_\alpha$, 
which amounts to simultaneously describing linearized deformations of
all minimal branes $\cM_{\bf m}$ with a fixed value of $\ell=-1+\sum_{i=1}^s{m_i}$.
This is especially convenient since \rf{Jdef} has the same form one would
encounter for linearized boundary deformations of a minimal brane at the
minimal model point, so we can use this `minimal model' form even though we
are studying deformations of minimal branes at a point in the closed
string moduli space which is away from $t=0$.
Implementing obstructions to \rf{Jdef} will give 
a 'total' deformation space ${\cal S}^{(\ell+1)}_t$ consisting of a
finite number of points, and one
must use the supplementary information provided by ${\bf m}$ in order to
identify the point associated with a given minimal brane.

For a linearized deformation \rf{Jdef}, we can extend $E$ to a non-polynomial
function defined through:
\beql{Edef}
E(x;t,u)=\frac{W(x;t)}{J(x;u)}~~.
\eeql 
Then the condition that $E$ be a polynomial in $x$ (i.e. $E_-=0$, where the subscript
denotes the singular part) imposes nonlinear
constraints on $u_\alpha$, which recover the `total' deformation space:
\beq
{\cal S}^{(\ell+1)}_t:=\{u\in \C^{\ell+1}|E_-(x;t,u)=0\}~~.
\eeq 
This consists of a finite number of points, corresponding to a choice
of $\ell+1$ zeroes of $W(x)$ in order to make up the polynomial
$J^{(\ell+1)}(x)$; of course, each such choice corresponds to a given
minimal brane $\cM_{\bf m}$. 

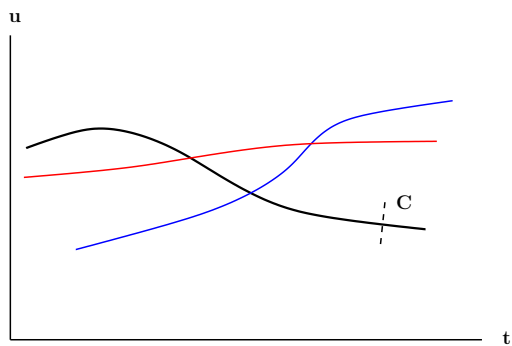
\begin{figure}[hbtp]
\begin{center}
\scalebox{0.3}{\input{cardy_moduli.pstex_t}}
\end{center}
     \caption{Schematic depiction of the 
       `total' joint deformation space ${\cal Z}^{(\ell+1)}$. Each
       branch is associated with a minimal brane, and carries a
       $\C^*$ bundle corresponding to the constant $C$ in
       (\ref{minimal_gen}). 
%       (strictly speaking, the modulus $C$ is not
%       physically relevant, as explained in Section 7) 
}
  \label{minimal_moduli}
\end{figure}

If we let the bulk moduli vary as well, we find the `total'
joint deformation space:
\beq
\label{tot_joint}
{\cal Z}^{(\ell+1)}=
\{(t,u)\in \C^{\ell+1}\times \C^{k+1}|E_-(x;t,u)=0\}~~,
\eeq
which was computed for some examples in \cite{BHLS}.  This affine algebraic variety is a
branched multicover the affine space $\C^{k+1}$ of
closed string moduli. The branching divisor coincides with the
discriminant locus of $W(x)$.  The joint deformation
space of each $\cM_{\bf m}$ (with $\sum_{i=1}^s{m_i}=\ell+1$) is a
$\C^*$-bundle over the corresponding branch of ${\cal Z}^{(\ell+1)}$
(figure \ref{minimal_moduli}).

%%%%%%%%%%%%%%%%%%%%%%%%%%
\subsection{Tachyon condensation and $D$-brane composites}
\label{tachcond}
%%%%%%%%%%%%%%%%%%%%%%%%%%

We next consider deformations of a tachyon composite formed 
from a system of two branes $\cM$ and $\cN$. After explaining how 
the composite can be identified with an object of $\DG_W$, we will
extract the appropriate parameterization of its linearized
deformations. 

We start by discussing when a pair of odd
morphisms in $\DG_W$ defines an allowed tachyon condensate, thus
leading to a D-brane composite. 
Consider two objects $\cM,\cN$ and odd morphisms $f:\cM\rightarrow \cN$, $g:\cN\rightarrow \cM$
in $\DG_W$, with components $f_0:\Mp\rightarrow \Nm, f_1:\Mm\rightarrow \Np$
and $g_0:\Np\rightarrow \Mm, g_1:\Nm\rightarrow \Mp$. We want to know
the conditions under which the maps $f, g$ can be interpreted as 
(topological) tachyon vevs arising from strings
stretching between the branes $\cM,\cN$. In that case, one obtains a D-brane
composite "glued" by the tachyon vevs $f,g$ and this physical interpretation 
requires that the morphism pair 
$\cM{\scriptsize \begin{array}{c} f \\ \rightleftarrows \\ g\end{array}}\cN$ 
be identified with a D-brane of our Landau-Ginzburg theory, i.e. 
an object of $\DG_W$. To understand the identification, let us decompose the system
as shown below:
\beql{conedia}
\xymatrix@R+40pt{
\cM\ar@<-1.0ex>[d]_{f}
\ &\ar@{}[d]_{{\textstyle\cong\ }}\ \ \ &\ 
\Big(
\Mm \ 
\ar@<-1.2ex>[dr]_{\!\!\!\! f_0~}
~
\ar@<0.6ex>[r]^{E_M} 
&
\ \Mp 
\ar@<-1.0ex>[dl]^{\!\!\!\!\!\!\!\!\!\!\!\!\!\! f_1~~~~~~}
\ar@<0.6ex>[l]^{J_M}
\Big)
\\
\cN\ \ar[u]_{g} \ &\ \ \ &\ \Big(
\Nm \ 
\ar[ur]_{g_0}
\ar@<0.6ex>[r]^{E_N} 
&
\ \Np
\ar@<0.6ex>[l]^{J_N}
\ar[ul]_{g_1}
\Big)
}
\eeql
Since $(\Mp,\Mm)$ and $(\Np,\Nm)$ can be identified with $D2$ brane-antibrane pairs
of the associated sigma model, it is clear that all eight maps on the right
should be viewed as tachyon vevs of that model, arising from strings stretched
between the sigma-model
branes $\Mp,\Np$ and their antibranes $\Mm,\Nm$. The net brane content of this
system is given by the module $P_0:=\Mp\oplus \Np$, while the net antibrane
content is $P_1:=\Mm\oplus \Nm$. Moreover, the net tachyon vevs are described by
the odd morphisms $J:P_0\rightarrow P_1$ and $E:P_1\rightarrow P_0$ given by
combining the relevant contributions:
\beq
\label{EFcomp}
J:=\left[\begin{array}{cc}J_M& g_0\\ f_0 & J_N\end{array}\right]~~,~~
E:=\left[\begin{array}{cc}E_M& g_1\\ f_1 & E_N\end{array}\right]~~.
\eeq
In these block matrices of morphisms, the column blocks are ordered by
$\Mp,\Np$ (in $J$) and $\Mm, \Nm$ (in $E$), while the row blocks are ordered
as $\Mm, \Nm$ (in $J$) and $\Mp,\Np$ (in $E$). 
The total tachyon condensate in the new object 
$\cP=
(P_0 {\scriptsize \begin{array}{c} E \\ \rightleftarrows \\ J\end{array}} P_1)$
is:
\beq
\label{Dcomp}
D=\left[\begin{array}{cc}0& E\\ J &
    0\end{array}\right]~~.
\eeq
For the interpretation as a composite to hold, this must satisfy the condition
$D^2=W\bfone \Longleftrightarrow EJ=JE=W\bfone$, 
in which case $\cP$ is an object of $\DG_W$ with which the
morphism pair $\cM{\scriptsize \begin{array}{c} f \\ \rightleftarrows \\
    g\end{array}}\cN$ should be identified. Using expressions (\ref{EFcomp}), 
we find that this condition is equivalent with the following two systems of constraints: 
\begin{eqnarray}
\label{gc1}
J_ME_M+g_0 f_1&=&W\bfone\nn\\
J_NE_N+f_0 g_1&=&W\bfone\nn\\
J_M g_1+g_0 E_N&=&0\\
J_N f_1+f_0 E_M&=&0\nn~~
\end{eqnarray}
and:
\begin{eqnarray}
\label{gc2}
E_MJ_M+g_1 f_0&=&W\bfone\nn\\
E_NJ_N+f_1 g_0&=&W\bfone\nn\\
E_M g_0+g_1 J_N&=&0\\
E_N f_0+f_1 J_M&=&0\nn~~.
\end{eqnarray}
A pair $\cM{\scriptsize \begin{array}{c} f \\ \rightleftarrows \\
    g\end{array}}\cN$ in $\DG_W$ which satisfies (\ref{gc1}) and (\ref{gc2}) will
    be called a {\em two-term generalized complex} (the terminology
    follows \cite{CIL2}). Our discussion
    shows that: 

\

\noindent{\em Two term generalized complexes over $\DG_W$ can be identified with objects of
    $\DG_W$.} 

\

It can be shown that this identification is compatible with the category
structure, namely such generalized complexes form a differential
graded (dG) category which is
equivalent with a full sub-category of $\DG_W$.

Thus two-term generalized complexes describe D-brane composites
obtained by tachyon condensation in a D-brane pair of the Landau-Ginzburg
    model. That condensation processes lead back to objects of
    the original D-brane category $\DG_W$ reflects the fact that this
    dG-category is quasiunitary in the sense of \cite{CIL2,CIL3} (namely, the
    original D-brane category is large enough to model the result of any
    D-brane composite formation process). In fact, the argument
    presented above is very similar to that given in \cite{CIL2} for the case
    of critical string theories. In particular, it is easy to extend this
    argument to generalized complexes built from an arbitrary number of
    objects of $\DG_W$, and prove that the obvious differential graded category defined by 
    such objects is equivalent with $\DG_W$. We stress that this
    approach is both physically and mathematically 
    fundamental and should precede the `on-shell' approach
    based on the cone construction, as explained in detail in
    \cite{CIL2} and in the review \cite{CIL5}. The fundamental approach is
    off-shell since this includes {\em dynamical} information about
    the underlying theory. Mathematically, triangulated categories
    do not suffice, due to the the well-known lack of
    naturality of distinguished triangles. It is by now
    well-understood that such categories should be promoted to
    differential graded or $A_\infty$ categories in order to avoid
    this problem. This, of course, has a clear physical meaning as
    explained in detail in \cite{CIL2,CIL3,CIL4,CIL5,CIL6,CIL7,CIL8,CIL9,CIL10}. 
    The most clear-cut example is the case of A-models on
    Calabi-Yau manifolds, for which the derived category of the Fukaya
    category cannot even be defined within another approach. 

We next discuss the spectrum and linearized deformations of the
composite $\cM{\scriptsize \begin{array}{c} f \\ \rightleftarrows \\
    g\end{array}}\cN$. Suppose that we are given maps $f,g$ satisfying
conditions (\ref{gc1}) and
(\ref{gc2}). Then a variation $(\delta E_M, \delta J_M, \delta E_N, \delta
J_N, \delta f, \delta g)$ agrees with these constraints
if and only if the following Maurer-Cartan equation is satisfied: 
\beq
(\delta D)^2+[D,\delta D]=0~~,
\eeq
where: 
\beq
\delta D=\left[\begin{array}{cc}0& \delta E\\ \delta J &
    0\end{array}\right]~~
\eeq
is the induced variation of the tachyon condensate. 
The odd spectrum of the composite 
$P\equiv (\cM{\scriptsize \begin{array}{c} f \\ \rightleftarrows \\
    g\end{array}}\cN)$ is given by solutions of the linearized equation: 
\beq
[D,\delta D]=0~~,
\eeq
modulo states of the form $[D,A]$ with even $A$ --- this, of course,
recovers the odd cohomology $H_D^1(\End_{\C[x]}(P))=H_{D}^1(\End_{\DG_W}(\cM\oplus \cN))$. 

An important case, which will be relevant below, arises for $f,g=0$,
when the total tachyon condensate takes the form:
\beq
D=D_M\oplus D_N=
\left[\begin{array}{cccc}0& 0&E_M&0\\0&0&0&E_N\\ J_M&0&0&0\\ 0&J_N&0&0\end{array}\right]~~,
\eeq
which amounts to $E=E_M\oplus E_N$ and $J=J_M\oplus J_N$.
This corresponds to starting with the direct sum object $\cM{\scriptsize
\begin{array}{c} 0 \\ \rightleftarrows \\
    0\end{array}}\cN\equiv \cM\oplus \cN$~. Then the
spectrum of $\cP$ is the direct sum
$\oplus_{A,B=M,N}{H_{D_{AB}}(\Hom_{\C[x]}(A,B))}$, i.e. the total spectrum
of boundary and boundary condition changing observables in the
system of independent branes $\cM$ and $\cN$. This implies that we
can parameterize linear deformations through:
\beq
\label{deltaDcomp}
\delta D=\sum_\alpha{\eta_\alpha^{M}\psi_\alpha^M}+
\sum_\beta{\eta_\beta^{N}\psi_\beta^N}+
\sum_\gamma{\eta_\gamma^{MN}\psi_\gamma^{MN}}+
\sum_\gamma{\eta_\gamma^{NM}\psi_\gamma^{NM}}~~,
\eeq
where $\eta$ are even coordinates  while $\psi^{AB}$ form a basis
of $H^1(\Hom(A,B))$, with $\psi^{A}:=\psi^{AA}$.
 Since $D$ describes the tachyon condensate of the brane $\cP \in Ob
\DG_W$,  the arguments of Subsection \ref{LGB} imply
that the associated variation of the boundary action is linear in
all deformation parameters:
\begin{eqnarray}
\label{Sbcdef}
\delta S_\del &=&
\sum_\alpha{\eta_\alpha^{M}\int_{\partial \Sigma}{dx~\{G^{b,M}_{-1},\psi^M_\alpha\}}}+
\sum_\beta{\eta_\beta^{N}\int_{\partial \Sigma}{dx~\{G^{b,N}_{-1},\psi^N_\beta\}}}\nn\\
&+&\sum_\gamma{\eta_\gamma^{MN}\int_{\partial \Sigma}{dx~\{G^{b,MN}_{-1},\psi^{MN}_\gamma\}}}+
\sum_\gamma{\eta_\gamma^{NM}\int_{\partial \Sigma}{dx~\{G^{b,NM}_{-1},\psi^{NM}_\gamma\}}}~~.
\end{eqnarray}

%%%%%%%%%%%%%%%%%%%%%%%%%%%%%%%%%%%%%%%%%%%
\section{A closed form for the effective potential}
%%%%%%%%%%%%%%%%%%%%%%%%%%%%%%%%%%%%%%%%%%%

%%%%%%%%%%%%%%%%%%%%%%%%%%%%%%%%%%%%%%%%%%%
\subsection{The effective potential for minimal branes}
%%%%%%%%%%%%%%%%%%%%%%%%%%%%%%%%%%%%%%%%%%%

In \cite{HLL}, we derived the consistency conditions for open-closed
topological string amplitudes on the disk (namely the open string
version of the WDVV equations, which includes the extension of 
the $A_\infty$ relations of \cite{Gaberdiel} to non-critical
strings).  As shown there, appropriately symmetrized open string
correlators integrate to the generating function $\weff$. In our case,
this potential is defined on the total space $\C^{k+1}\times \C^{\ell+1}$
of joint linearized deformations, which plays the role of ambient
space for the total joint deformation space ${\cal Z}^{(\ell+1)}$
of equation (\ref{tot_joint}).
 
By generalizing results for minimal branes in various minimal models
with low $k$ and $\ell$, 
the following closed expression for the effective potential was
obtained in \cite{HLL}:
\beql{weffdef}
\weff(t,u)\ =\ \sum_{m=0}^{k+2} g_{k+2-m}(t)h_{m+1}\lab\ell(u)~~.
\eeql
Here $g_{k+2-m}(t)$ are the coefficients of the bulk Landau-Ginzburg superpotential
\rf{wlg} and $h_{m+1}\lab\ell(u)$ are the symmetric polynomials defined through
the expansion:
\beql{genfct}
\log\Big[1-\sum_{n=1}^{\ell+1} u_n y^n\Big] \ :=\ 
\sum_{m=1}^\infty h_m\lab\ell(u) y^m\ .
\eeql 
That \rf{weffdef} is indeed correct for all $k,\ell$ is a
conjecture still awaiting proof, and we intend to address this in
future work. In this paper, we shall
accept that \rf{weffdef} is generally valid and investigate
its consequences.

The deeper meaning of \rf{weffdef} is that correlators involving
only boundary fields are completely determined by combinatorics.
The property which underlies the appearance of symmetric functions
is that all non-trivial correlators of odd\footnote{While correlators
involving even boundary observables are typically non-zero as well, the
corresponding terms in the effective potential drop out upon
(super-)symmetrization, so they do not play a role for our purpose.
However, they must be taken into account when solving the consistency
constraints \cite{HLL}.} boundary observables have the same value.
More precisely, they have the same non-zero value whenever the
charge superselection rule is satisfied. With our normalization
conventions, such nontrivial correlators are given by:
\beql{bpcorr}
\langle\,\psi_{\al_1}\,\psi_{\al_2}\,\int G^-\psi_{\al_3}\,\dots\, 
\int G^-\psi_{\al_{n-1}}\,\psi_{\al_n}\,\rangle\ = -\frac1{k+2}\ .
\eeql
The $n-3$ integrations complicate the direct evaluation of these
correlators due to the presence of contact terms, which is why we
resorted to determining them by solving the consistency conditions.

It is possible to cast \rf{weffdef} into a more elegant form.
For this, notice that the substitution $y=1/x$ reduces \rf{genfct} to:
\beql{Jexpan}
\log J(x;u) = (l+1)\log x+\sum_{m=1}^\infty h_m\, x^{-m}~~,
\eeql
where $J(\vp,u)$ is the boundary superpotential, 
parametrized as in \rf{Jdef} (the expansion in \rf{Jexpan} is valid for large $x$).
The effective potential \rf{weffdef} can thus be written as:
\beql{Weff}
\weff(t,u)=-\oint_{\cal C}{\frac{dx}{2\pi i}\wlg(x;t)\log J(x;u)}~~,
\eeql
where ${\cal C}$ is a closed counterclockwise contour encircling all
$D0$-branes (i.e. all zeroes $x_i(u)$ of $J(x)$) and all cuts of the
logarithm. 
Relation \rf{Weff} is ambiguous due to the
need of choosing appropriate branch cuts, but the ambiguity amounts to the
freedom of adding an inessential constant to $\weff$. 

>From the interpretation of $\weff$ as a deformation potential 
\cite{HLL,CIL4}, 
we expect that its $u$-critical set, defined by\footnote{We
treat bulk deformations as non-dynamical background fields,
which is warranted at weak string coupling.}:
\beq
\label{Zeff}
{\cal Z}_{crit}=\{(t,u)|\partial_u\weff(t,u)=0\}~~
\eeq
should agree {\em locally} with the total joint deformation locus
(\ref{tot_joint}). More precisely, ${\cal Z}_{crit}$ should coincide with a
branch of (\ref{tot_joint}), provided that we restrict both to a small enough
vicinity of a point $(t,u)$ which lies on such a branch. 
Thus we are
interested in polynomials $J(x;t,u)$ as in (\ref{Jdef0}) which are close to a polynomial
$J_0(x;t)$ associated with a minimal brane $\cM_{\bf m}$
for which $\sum_{i=0}^s{m_i}=\ell+1$ (in particular,  $(\frac{W}{J_0})_-=0$). 

This expectation is in fact easy to check by
writing $J(x;u)=\prod_{i=0}^{\ell}{(x-x_i(u))}$, which implies: 
\beql{eqom}
\partial_{u_\alpha}\weff(t,u)=\oint_{\cal C}{\frac{dx}{2\pi i}
\left[W(x;t)\sum_{i=0}^\ell{\frac{\partial_{u_\alpha}x_i(u)}{x-x_i(u)}}
\right]}=\sum_{i=0}^\ell{W(x_i(u);t)\partial_{u_\alpha}x_i(u)}~~.
\eeql
Thus the $u$-critical set of $\weff$ is described by the linear
system:
\beq
\label{linsys}
\sum_{i=0}^{\ell}{\partial_{u_\alpha}x_i(u)W(x_i(u);t)}=0
\eeq
for the $\ell+1$ unknowns $W(x_i(u))$. 
Now notice that the $\ell+1$ parameters $u$ in \rf{Jdef} 
suffice to specify the roots of the monic degree
$\ell+1$ polynomial $J(x)$. As a consequence, the discriminant:
\beq
\label{Delta}
\Delta(u):=\det(\partial_{u_\alpha} x_i(u))
\eeq
is generically non-vanishing. 
Hence the only solution of (\ref{linsys}) is $W(x_i(u))=0$ for all
$i=0\ldots \ell$. Thus each root of the polynomial
$J(x)$ is also a root of $W(x)$. Since $J$ is close to $J_0$, which divides
$W$, the only possibility is that the multiplicities of the roots 
are smaller in $J(x)$ than in $W$. Thus $J$ must divide $W$, and 
${\cal Z}_{crit}$ must coincide with the $J_0$-branch of
${\cal Z}^{(l+1)}$ when restricted to a small enough vicinity of $J_0$. 

Notice that this is a purely
local statement. The variety ${\cal Z}_{crit}$ contains components associated
with polynomials $J$ that do not divide $W$. However, such components do
not intersect the factorization locus ${\cal Z}^{(\ell+1)}$, so agreement is
guaranteed in the vicinity of any true solution of the factorization problem 
(which, of course, is all that can be expected from the local analysis of
\cite{HLL,CIL4}). 

Although the factorization $W = J E$ persists along the $u$-critical
set $\cZ_{crit}$, the cohomology in the boundary sector
may change along this locus. In the remainder of this subsection, 
we discuss the condition on $\weff(t,u)$ which 
ensures the preservation of a non-trivial spectrum.  On this account
we differentiate equation (\ref{eqom}) a second time and obtain:
\beql{secderWeff}
  \partial_{u_\alpha}\partial_{u_\beta}\weff(t,u) =
  \sum_{i=0}^\ell {W(x_i(u);t)\ 
    \left(\partial_{u_\alpha} \partial_{u_\beta} x_i\right)} +
  \sum_{i=0}^\ell {\partial_{x_i}W(x_i(u);t)\ 
    \left(\partial_{u_\alpha} x_i\partial_{u_\beta} x_i\right)}\ 
~~.
\eeql
Suppose we stay at a point on the factorization locus, and we require,
in addition, that $J | E$, i.e., that we are on the sub-locus:
\beql{specloc}
  \cZ_{spec} := \{ (t,u) ~|~ W = J E,~J | E\} \subset \cZ_{crit}
\eeql
Since the boundary preserving spectrum is governed by the ideal
${\cal I}$ as defined in (\ref{boundideal}), this ensures that
the number of odd (and even) cohomology classes takes the maximal value,
$\ell+1$. Note that $\cZ_{spec}$ can equivalently be described by 
$\cZ_{spec} = \{ (t,u)~|~ J | W,~J| W'\}$. Therefore, we see from
(\ref{secderWeff}) that
\beql{speccond}
  \partial_{u_\alpha}\weff(t,u) =
  \partial_{u_\alpha}\partial_{u_\beta}\weff(t,u) = 0
  \quad\mathrm{on}~\cZ_{spec}~~.
\eeql
In order to show that (\ref{speccond}) is true only on
$\cZ_{spec}$, we look at the vicinity of a point
$(t_0,u_0) \in\cZ_{spec}$, with $J_0$ and $E_0=hJ_0$. Then, by the
same line of argumentation as above, the non-vanishing discriminant
$\Delta(u)$ ensures that 
\beql{potspecloc}
  \cZ_{spec} = \{ (t,u) ~|~ \partial_{u_\alpha}\weff(t,u) =
  \partial_{u_\alpha}\partial_{u_\beta}\weff(t,u) = 0\}~~.
\eeql
An analogous argument can be made for the situation where $J$ and
$E$ share a common factor $G$, whose degree is smaller than of $J$
(cf. \rf{JpG}). Then only a corresponding subset of the cohomology
survives, and this reflected in a increased rank of
$\partial_{u_\alpha}\partial_{u_\beta}\weff$.

In physical terms this finding can be interpreted as follows: On
the factorization locus $\cZ_{crit}$ where $W=JE$, the boundary preserving
parameters $u_\alpha$ do not have tadpoles and thus the theory has
a stable, supersymmetric vacuum; however a non-trivial spectrum of
boundary operators is not guaranteed. Only on the sub-locus
$\cZ_{spec}\subset \cZ_{crit}$ one has a non-trivial
spectrum, and this is reflected in zero eigenvalues of the
`mass-matrix' $\partial_{u_\alpha}\partial_{u_\beta}\weff$. 

%%%%%%%%%%%%%%%%%%%%%%%%%
\subsection{The effective potential for general B-type branes}
%%%%%%%%%%%%%%%%%%%%%%%%%

By solving the consistency conditions
of \cite{HLL} for correlators of various low-$k$ models with several
minimal branes, we checked that the
equality of correlators \rf{bpcorr} also holds for amplitudes
involving odd boundary condition changing observables which mediate
between such 
branes.  Assuming that this property holds in general
and accounting for the relevant combinatorics, we are lead to the
following expression for the disk generating function of strings
ending on arbitrary branes $\cM=(M,D)$:
\beql{weff}
\weff(t,\eta)\ =-\oint_{\cal C}{\frac{dx}{2\pi i} 
\log( \det J(x;\eta)\big) \wlg(x;t)}~~.
\eeql
This amounts to replacing $J$ by $\det J$ in \rf{Weff}. 

The proof of local agreement of the critical locus of $\weff$
with the deformation space is similar to the case of minimal branes.
Let us parameterize deformations of the $r\times r$ matrix
$J$ by $\eta=(\eta_\alpha)$ with $\alpha=0\dots H$, and write 
$\det J(x;\eta)=\prod_{i=0}^L{(x-x_i(\eta))}$ as a monic polynomial in $x$,
where $L+1$ is the degree of $\det J(x)$. 
We also assume that $H\geq L$ and that the $H+1$ by $L+1$ matrix:
\beq
A(\eta):=(\partial_\alpha x_i(\eta))~~.
\eeq
has maximal rank. Then:
\beq
\partial_{\eta_\alpha}\weff(t, \eta)=\oint_{\cal C}{\frac{dx}{2\pi i}
\left[W(x;t)\sum_{i=0}^L{\frac{\partial_{\eta_\alpha}x_i(\eta)}{x-x_i(\eta)}}
\right]}=\sum_{i=0}^L{W(x_i(\eta);t)\partial_{\eta_\alpha}x_i(\eta)}~~
\eeq
and we find that the $\eta$-critical locus ${\cal Z}_{crit}$ of ${\cal
  W}_{eff}$ is characterized by the condition that all roots of $\det J$ must
also be roots of $W$. Since this is obviously the case along the joint deformation
space ${\cal Z}$ (where $JE=W{\bf 1}$ implies $\det J|W^r$), the inclusion
${\cal Z}\subset {\cal Z}_{crit}$ is immediate. Local agreement of ${\cal
  Z}_{crit}$ with ${\cal Z}$ after
restriction to a sufficiently small vicinity of a point lying on
${\cal Z}$ follows by a simple continuity argument, as in the minimal case. 
We note that the inclusion ${\cal Z}\subset {\cal Z}_{crit}$  
also follows directly from \rf{weff}, which implies\footnote{
The sign change in the last equation reflects the fact that swapping $J$ and $E$
exchanges branes with antibranes.}:
\beql{factoriz}
\del_{\eta_\al}\weff(t,\eta)\ &=&\
-\oint_{\cal C}{\frac{dx}{2\pi i}\, \Tr [E(x;t,\eta)\del_{\eta_\al} J(x;\eta)]}\
=\oint_{\cal C}{\frac{dx}{2\pi i}\,\,
\Tr[J(x;\eta)\del_{\eta_\al} E(x;t,\eta)]}~~.\nn
\eeql
The right hand side of this identity vanishes along  ${\cal Z}$, since by definition
the matrix
$E(x;t,\eta):=\frac{W(x;t)}{J(x;\eta)}$ has no singular terms there.

Thus the boundary critical set of $\weff$ agrees with the matrix
factorization locus. This provides further evidence for our general ansatz
\rf{weff}.

%%%%%%%%%%%%%%%%%%%%%%%%%%%%%%%%%%%%%%%%%%%
\subsection{Interpretation through holomorphic matrix models}
%%%%%%%%%%%%%%%%%%%%%%%%%%%%%%%%%%%%%%%%%%%

Our proposal \rf{weff} for the effective potential admits
a matrix model interpretation. 
For this, consider an antiderivative $V(x)$ of $W(x)$, i.e. a
polynomial $V(x;t)$ in $x$ (whose coefficients depend parametrically on $t$)
which satisfies: 
\beq
\partial_x V(x;t)=W(x;t)~~
\eeq
(clearly $V$ is defined only up to addition of a function  $c(t)$ which
is independent of $x$).
Then integration by parts casts \rf{Weff} into the form:
%\beql{Weffmatrix}
$$
\weff(t,\eta)=\oint_{\cal C}\frac{dx}{2\pi i}
V(x;t)\sum_{i=0}^L\frac{1}{x-x_i(\eta)}=
\sum_{i=0}^L{V(x_i(\eta),t)}
%=\Tr V(X(\eta),t)
~~,
$$
%\eeql
where $\det J(x;\eta)=\prod_{i=0}^L{(x-x_i(\eta))}$ as before.
Viewing the zeros $x_i(\eta)$ of $\det J(x;\eta)$ as eigenvalues
of a complex $(L+1)\times (L+1)$ matrix $X(\eta)$, we can write the
effective potential as:
\beql{Weffmatrix}
\weff(t,\eta)=
%\oint_{\cal C}\frac{dx}{2\pi i}
%V(x;t)\sum_{i=0}^L\frac{1}{x-x_i(\eta)}=
%\sum_{i=0}^L{V(x_i(\eta),t)}=
\Tr~ V(X(\eta),t)~~.
\eeql
Thus $\weff$ coincides with
the classical action\footnote{In this paper, we
consider only the ``small
phase space''. It would be interesting to extend the correspondence
by coupling to topological gravity and including gravitational
descendants. Presumably this involves the full dynamics of the
holomorphic matrix model, rather than simply its classical action.}
of a holomorphic matrix model as defined and studied in
\cite{holo} (the matrix model is holomorphic rather than Hermitian 
because the eigenvalues $x_i(u)$ are complex). 
The zeroes $x_i(\eta)$ of $\det J$ can be viewed as the locations
of  $D0$-branes in the complex plane (=the target space of the
Landau-Ginzburg model). Equation \rf{Weffmatrix} shows that
$\weff$ is the `potential energy' of this system of $D0$ branes when the
latter is placed in the external `complex potential' $V(x)$. 
Each critical configuration of this $0$-brane system 
corresponds to a  deformation of the underlying Landau-Ginzburg
brane. 

It has been known for a long time that the
generalized Kontsevich model \cite{GKM} is closely related to closed string 
minimal models coupled to
topological gravity, but the open string version of this
correspondence is not well understood. A link
between certain topological $D$-branes and the auxiliary (Miwa) matrix of
the Kontsevich model was proposed in \cite{ADKMV}, in the context of
a non-compact Calabi-Yau realization of the underlying closed string model.

Our Landau-Ginzburg description gives a direct relation, which differs in spirit
from that proposed in \cite{ADKMV}:  in the presence of
several D-branes, the bulk Landau-Ginzburg field $x$ is effectively promoted to
a matrix $X(\eta)$. In \cite{ADKMV}, D-brane positions were mapped to the {\em
auxiliary} matrix of the generalized Kontsevich model, so they
parameterize backgrounds for the model's dynamics. In our case,
D-brane positions are truly dynamical, being encoded by the matrix
{\em variable} itself. The reason for this difference is that we
study the D-brane potential (i.e. the generating function of
scattering amplitudes for strings stretching between D-branes)
rather than the flux superpotential (the contribution from RR flux
couplings to the closed string sector) considered in
\cite{ADKMV}.\footnote{Since D-branes carry RR charges,
they can be viewed as backgrounds inducing a flux superpotential,
which explains the different point of view used in \cite{ADKMV}.
Our interest, however, is in D-brane dynamics as dictated by
tree-level scattering amplitudes of open strings.}

%%%%%%%%%%%%%%%%%%%%%%%%%%%%%%%%%%
\section{Examples of moduli spaces and tachyon condensation}
\label{examples}
%%%%%%%%%%%%%%%%%%%%%%%%%%%%%%%%%%

In this section, we will apply our methods to analyze tachyon condensation
in a few examples, for which we will explicitly determine the moduli spaces and
their stratification.  Some examples of boundary flows\footnote{This
is somewhat loose language, since, strictly speaking, there are no
RG flows in the standard sense in topological models. More precisely,
one can define a sort of RG flow at the level of string field theory, but such flows are homotopy
equivalences with respect to the BRST operator so 
they always reduce to isomorphisms on-shell.} in minimal models were previously
discussed in \cite{freden,cappelli,Horiflows}.

Specifically, we study pure boundary deformations and switch off any bulk
moduli; thus we shall set $\wlg(x;t)=\frac1{k+2}x^{k+2}$.
Starting with a system  
$\cM_{\ell_1}\oplus \cM_{\ell_2}$ of two independent minimal branes, we are interested in D-brane
composite formation induced by turning on vevs for odd boundary
preserving and boundary changing observables.  This is the situation
considered in Subsection \ref{tachcond}, with the choice
$\cM=\cM_{\ell_1}$ and $\cN=\cM_{\ell_2}$. The effective potential
\rf{weff} takes the form: 
\beq
\label{weff_k}
\weff(u)=\frac{1}{k+2}\log\left[\frac{\det J(x;u)}{x^{\deg J(x)}}\right]\Big|_{x^{-k-3}}~~,
\eeq
where we indicated that we take the coefficient of
$x^{-k-3}$ in the large $x$ expansion of $\det J$ (notice that we
divide out $x^{\deg J(x)}$ under the logarithm in order to have a
well-defined Laurent expansion around $x=\infty$). 
Above, $J(x;u)\equiv J^{\ell_1,\ell_2}(x;u)$ is the `total' 
map of equation (\ref{EFcomp}), which
arises upon representing the composite $(\cM_{\ell_1}{\scriptsize
\begin{array}{c} f \\ \rightleftarrows \\
g\end{array}}\cM_{\ell_2})$ as a single object $( P_0 {\scriptsize
\begin{array}{c} E \\ \rightleftarrows \\ J\end{array}} P_1)$  of
$\DG_W$.  Then relations (\ref{omega_bc}) and (\ref{deltaDcomp}) give
the following explicit parameterization:
\beql{JJdef}
J^{\ell_1,\ell_2}(x;u)\ 
&=&\ 
\left[\begin{array}{cc}
J\lab{\ell_1+1}(x;u^{[11]}) & g\lab{\frac12(\ell_1+\ell_2)+1}(x;u^{[12]})\\
f\lab{\frac12(\ell_1+\ell_2)+1}(x;u^{[21]}) & J\lab{\ell_2+1}(x;u^{[22]})
\end{array}\right]\nn
\\
\ &=&\ 
\left[\begin{array}{cc}
x^{\ell_1+1}-\sum_{\alpha=0}^{\ell_1}u_{\ell_1+1-\alpha}^{[11]}x^\alpha 
& 
-\sum_{\gamma=0}^{\ell_{12}}u_{\frac12(\ell_1+\ell_2)+1-\gamma}^{[12]}x^\gamma
 \\
-\sum_{\gamma=0}^{\ell_{21}}u_{\frac12(\ell_1+\ell_2)+1-\gamma}^{[21]}x^\gamma
 &
  x^{\ell_2+1}-\sum_{\alpha=0}^{\ell_2}u_{\ell_2+1-\al}^{[22]}x^\alpha
\end{array}
\right]~~.
\eeql
Upon substitution into (\ref{weff_k}), this agrees with 
the disk generating function found by solving the consistency
constraints of \cite{HLL}.  Here
$\ell_{12}=\ell_{21}=\min(\ell_1,\ell_2)$ and we traded the generic
deformation parameters $\eta$ for parameters $u$ indexed in an
manner which denotes their formal $U(1)$ charges (cf. equation
\rf{bccharges}). Moreover, we used superscripts to indicate the
formal $U(1)$ charges of $J,f$ and $g$.

It is clear that the factorization locus in $u$-space (the locus
where the factorization constraint (\ref{factcond}) is satisfied)
is determined by the equation 
\beql{simplelocus}
\det J(x;u)=x^{\ell_1+\ell_2+2}\ .
\eeql

A simple way to satisfy this relation is to take all parameters $u$ to
vanish except for $u^{[12]}_\al$ (or $u^{[21]}_\al$), in which case 
the generic Smith normal form of $J$ is $(1,x^{\ell_1+\ell_2+2})$
-- where `generic' means that $u^{[12]}_{\frac12(\ell_1+\ell_2)+1}\neq
0$ (or  $u^{[21]}_{\frac12(\ell_1+\ell_2)+1}\neq
0$ ). This corresponds to the minimal brane content $\cM_{-1}\oplus
\cM_{\ell_1+\ell_2+1}$, where $\cM_{-1}$ is trivial.  However, if
we switch on only a single $u\Lab{12}_{1+j}$ (or $u\Lab{21}_{1+j}$), where
$j\in\{\frac{|\ell_1-\ell_2|}{2} \ldots \min(\frac{\ell_1+\ell_2}{2},
k-\frac{\ell_1+\ell_2}{2})\}$, the resulting Smith normal form corresponds
to the process:
\beql{fusionflows}
\cM_{\ell_1}\oplus \cM_{\ell_2}\ \mathop{\longrightarrow}^{u\Lab{12}_{1+j}\not=0}\
\cM_{\hll+j+1}\oplus \cM_{\hll-j-1}~~.
\eeql
This reproduces a result given in \cite{Horiflows}.  Notice that upon
setting $\ell_2=k-\ell_1$ and $j=k/2$, relation \rf{fusionflows} 
includes brane-antibrane annihilation (in that case,  the right hand
side is the closed string vacuum since $\cM_{k+1}=\cM_{-1}[1]$ and
$\cM_{-1}$ is the trivial brane).
 
The complete deformation space is much more complicated since it
is not restricted to purely block upper or lower off-diagonal
deformations. This space is determined by condition
\rf{simplelocus}, which amounts to a system of quadratic equations for the
parameters $u$ in $J(x;u)$. In general, this is a complicated affine
algebraic variety, containing a multitude of strata associated with
different Smith normal forms. The highest dimensional
stratum corresponds to the Smith form $J=\diag(1,
x^{\ell_1+\ell_2+1})$ and gives the minimal brane content 
$\cM_{-1}\oplus \cM_{\ell_1+\ell_2+1}$, as discussed above. 
Lower dimensional strata are
obtained when the greatest common divisor $G_1(x)$ of the $1\times
1$ minors of $J(x)$ is non-constant, and can be described
by common factor conditions between the entries of $J(x)$.
Because the general analysis of strata is rather complicated, it will
not be presented here.  Instead, we will illustrate it with a 
few examples further below. 

Before doing so, let us  mention that our considerations
are not limited to two-brane systems. In fact, one can consider
multi-brane configurations described by tachyon profiles
$J^{\ell_1,...,\ell_N}$, for example systems of $N$ `elementary'
branes $\cM_0$ described by:
\beql{J00}
J^{0,0,\dots0}(x;u)\ =\ \left(\begin{array}{cccc}
x-u_1\Lab{11} & - u_1\Lab{12} & \dots & - u_1\Lab{1N}\\
- u_1\Lab{21} & x-u_1\Lab{22} & & \vdots\\
\vdots &  & \ddots & \\
- u_1\Lab{N1} & \dots & & x- u_1\Lab{NN}
\end{array}\right)\ .
\eeql
We expect that arbitrary composites can be obtained by switching
on suitable combinations of moduli. One may wonder what happens if
$\sum \ell_i$ becomes arbitrarily large - clearly composites
$\cM_\ell$ with $\ell\geq k+2$ do not exist.  In fact, the factorization
condition (\ref{factcond}) ensures that such branes cannot be formed;
in other words, there cannot exist flat directions in $\weff$ that
would lead to such branes. However, configurations with an arbitrary
number of allowed branes $\cM_\ell$ with $\ell< k+2$ can be obtained.

As a specific example, consider a system of $N$ identical branes $\cM_\ell$, with
$\ell\geq N-2$. One can check that switching on the tachyons
$\{u_*\}\equiv\{u\Lab{1,N}_{N-1},
u\Lab{2,N-1}_{N-2}...u\Lab{N/2,N/2+1}_1\}$ (for $N$ even) or
$\{u_*\}\equiv\{u\Lab{1,N}_{N-1},
u\Lab{2,N-1}_{N-2}...u\Lab{(N-1)/2,(N+1)/2+1}_2\}$ (for $N$ odd)
produces a Smith form corresponding to the following deformation:
\beql{SFfusionflows}
\Big(\,\cM_{\ell}\Big)^{\oplus N}\ \mathop{\longrightarrow}^{u_*\,\not=0}\
\bigoplus_{\ell'} C_{N+2,\ell}^{\qquad\ \ell'}\cM_{\ell'}\ .
\eeql
The quantities $C_{N+2,\ell}^{\qquad\ \ell'}$ are the $SU(2)$ fusion rule
coefficients at level $k$ (with the understanding that $\ell=i-2$
labels an $i$-dimensional representation of $SU(2)$). This reproduces
the boundary fusion rules found in \cite{freden}, which are based
on the coset construction of the ${\cal N}=2$ minimal models. It
would be interesting to see whether there is a deeper relationship
between matrix factorizations, Smith normal forms and boundary
fusion rings -- perhaps similar in spirit to~\cite{GepFR}.

After these general remarks, we now turn to the detailed analysis 
of a few examples. We will concentrate on deformations obtained by turning
on tachyon vevs in systems of two minimal branes.

%%%%%%%%%%%%%%%%%%%%%%%%%%%%%%%
\subsection{$\cM_{0}\oplus \cM_{\ell-1}\rightarrow \cM_{-1}\oplus
  \cM_{\ell}$  (where $\cM_{-1}$ is trivial).} 
%%%%%%%%%%%%%%%%%%%%%%%%%%%%%%%

\begin{figure}[hbtp]
\begin{center}
\scalebox{0.7}{\input{moduli1.pstex_t}}
\end{center}
     \caption{Schematic description of the moduli space for the composite of 
$\cM_0$ and $\cM_{\ell-1}$. }
  \label{moduli1}
\end{figure}
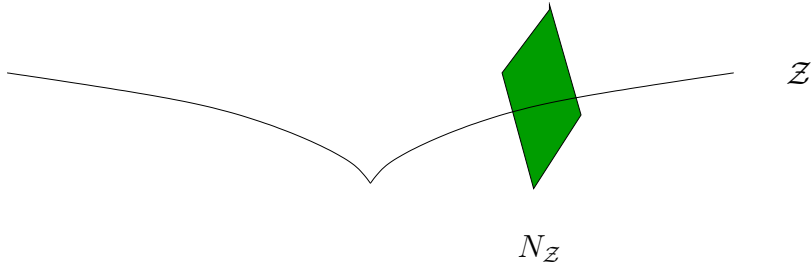

This is the basic mechanism for recursively building up any D-brane
$\cM_\ell$ from the `elementary' branes $\cM_0$. We start with
$$
J^{\ell_1=0,\ell_2=\ell-1} = \left[\begin{array}{ccc}
x-u_1\Lab{11} & -u_{\half(\ell+1)}\Lab{12}\\
-u_{\half(\ell+1)}\Lab{21} & x^{\ell}-\sum_{i=0}^{\ell-1} u_{\ell-i}\Lab{22}x^i\\
\end{array}\right]~~,
$$
which depends linearly on $\ell+3$ complex parameters $u$.
The determinant has the form: 
\beq
\det J=x^{\ell+1}-(u_1\Lab{11}+u_1\Lab{22})x^\ell+
\sum_{i=1}^{\ell-1}(u_1\Lab{11}u_{\ell-i}\Lab{22}-
u_{\ell+1-i}\Lab{22})x^i+u_1\Lab{11}u_{\ell}\Lab{22}
-u\Lab{12}_{\frac{\ell+1}{2}}u\Lab{21}_{\frac{\ell+1}{2}}~~,
\eeq
and gives a complicated expression for $\weff$. The
critical locus  ${\cal Z}_{crit}$ is
characterized by the condition $\det J=x^{\ell+1}$, which gives the system of
equations:
\beql{eqs}
u_1\Lab{11}+u_1\Lab{22}&=&0\non\\
u_1\Lab{11}u_{\ell-i}\Lab{22}-
u_{\ell+1-i}\Lab{22}&=&0~~,~~{\rm for}~i=1\ldots \ell-1\\
u_1\Lab{11}u_{\ell}\Lab{22}
-u\Lab{12}_{\frac{\ell+1}{2}}u\Lab{21}_{\frac{\ell+1}{2}}&=&0~~.\nn
\eeql
This can be solved recursively in terms of $a:= u_1\Lab{11}$. From
the first equation we find $u_1\Lab{22}=-a$, while the $\ell-1$
conditions in the middle give the recursion relations: 
\beq
u_{i+1}\Lab{22}=a u_i\Lab{22}~~{\rm~for~}~~i=1\ldots \ell-1~~,
\eeq
with the solution:
\beq
u_{i}\Lab{22}=-a^i~~{\rm~for~}~~i=1\ldots \ell~~.
\eeq
Substituting in the final relation of \rf{eqs}, we obtain: 
\beq
\label{Zex1}
{\cal Z}~~:~~a^{\ell+1}+u
\Lab{12}_{\frac{\ell+1}{2}}u\Lab{21}_{\frac{\ell+1}{2}}=0~~.
\eeq
Thus the factorization locus ${\cal Z}$ is the affine 
complex surface defined by equation (\ref{Zex1}) in $\C^3$, which is the
well-known $A_\ell$ singularity. The singular point sits at the
origin $u=0$ of the parameter space, as expected from the presence of
obstructions to linearized deformations at that point. Equations \rf{eqs}
realize ${\cal Z}$ as a complete intersection in the original parameter space
$\C^{\ell+3}$. When moving along ${\cal Z}$, one
turns on vevs for tachyon fields between $\cM_{0}$ and 
$\cM_{\ell-1}$, thus forming a D-brane composite. Part of the virtual 
(i.e. linearized) deformations of this composite span the normal space to ${\cal
  Z}$ inside $\C^{\ell+3}$. Such normal deformations are of course obstructed, since 
the unobstructed directions are those tangent to ${\cal Z}$ (figure \ref{moduli1}). 
To identify the strata and minimal brane decompositions, it is convenient to introduce
the simplified notation:
\beq
u_1^{[11]}=a~~,~~u_{\frac{1}{2}(\ell+1)}^{[12]}=b~~,~~u_{\frac{1}{2}(\ell+1)}^{[21]}=c~~.
\eeq
Then $J$ takes the following form along the critical locus: 
\beq
J=\left[\begin{array}{ccc}
x-a & -b\\
-c& x^{\ell}-\sum_{i=0}^{\ell-1}{a^{\ell-i}x^i}\\\end{array}\right]~~,
\eeq
where the parameters are subject to the constraint:
\beq
\label{abc}
bc+a^{\ell+1}=0~~.
\eeq
If $a\neq 0$, this equation shows that $b\neq 0$ and $c\neq 0$, which implies: 
\beq
G_1=\gcd(x-a, -b, -c,x^{\ell}-\sum_{i=0}^{\ell-1}
{a^{\ell-i}x^i} )=1~~.
\eeq
Thus $p_1=1$, $p_2=\det J=x^{\ell+1}$ and $J$ can be brought to the form:
\beq
J\sim J_0=\left[\begin{array}{cc}1&0\\0&x^{\ell+1}\end{array}\right]
\eeq
by a double similarity transformation. In this case, we find the minimal brane
content $\cM_{-1}\oplus \cM_\ell$. The same situation occurs for $a=0$ with
$b\neq0$ or $a\neq0$. 

At the origin $a=0=b=c=0$ (the singular point of ${\cal Z}$), we find 
$G_1=\gcd(x,0,0,x^\ell)=x$, so $p_1=x$, $p_2=x^{\ell-1}$ and $J$ can be
brought to the form:
\beq
J\sim J_0=\left[\begin{array}{cc}x&0\\0&x^\ell\end{array}\right]~~.
\eeq
This gives the minimal brane content $\cM_{0}\oplus \cM_{\ell-1}$ which, as expected, is the
original D-brane system.

We conclude that the deformation space ${\cal S}={\cal Z}$ has two
strata, characterized by the integer two-vector $m=(m_1,m_2)$:
\beq
{\cal S}={\cal Z}={\cal O}_{1,\ell}\sqcup {\cal O}_{0,\ell+1}~~.
\eeq
Namely ${\cal O}_{1,\ell}$ is of the origin of the parameter space (the
singular point of the $A_1$ singularity ${\cal Z}$), while
${\cal O}_{0,\ell+1}$ is the complement of ${\cal O}_{1,\ell}$ inside ${\cal Z}$.
When moving away from the origin (even infinitesimally!),
the minimal brane content jumps from $\cM_{0}\oplus
\cM_{\ell-1}$ to $\cM_{-1}\oplus \cM_{\ell}$.

To find the potential along the space $N_{\cal Z}$ normal to ${\cal Z}$, 
notice that normal directions can be described by variables $s_1\dots s_{\ell+1}$ 
defined through the relations:
\begin{eqnarray}
%u_1\lab{11} &=& a~,\\
u_1^{[22]} &=& s_1 - a~,\nonumber\\
u_2^{[22]} &=& s_2 + s_1 u_1- a^2,\\
&\vdots&\nonumber\\
u_\ell^{[22]} &=& s_\ell + \sum_{j<\ell} s_j a^{i-\ell}- a^\ell~~,\nonumber
\end{eqnarray}
and:        
$$
u_{\half(\ell+1)}^{[12]}u_{\half(\ell+1)}^{[21]} =
s_{\ell+1}+ \sum_{j<\ell+1} s_j a^{\ell+1-j}- a^{\ell+1}.$$
With these substitutions, the effective potential \rf{weff} reduces
to that of the brane $\cM_\ell$, while $a$ and the mode corresponding 
to the ratio $u_{\half(\ell+1)}^{[12]}/u_{\half(\ell+1)}^{[21]}$ decouple:
$$
\weff^{0,\ell-1}(u)\ \rightarrow\ \log[x^{\ell+1}-\sum_{i=0}^\ell s_{\ell+1 -i}x^i]
\Big |_{x^{-k-3}}=\weff^{\ell}(s)~~.
$$
As expected, the
variables  $s_i$ can be identified with the 'special' deformation parameters
of $\cM_\ell$. The $s_i$-deformations are completely obstructed, as
shown by their appearance in the effective potential, and by the fact
that they spoil matrix factorization. This is in contrast with the flat
directions tangent to ${\cal Z}$, for which the factorization condition
is preserved. This can be checked by using the expressions:
\begin{eqnarray}
J &=& \left[\begin{array}{ccc}
x-a & -u_{\half(\ell+1)}\Lab{12}\\
-u_{\half(\ell+1)}\Lab{21} & x^{\ell-1}+\sum a^{\ell-1-i}x^i\\
\end{array}\right]~~,
\\
E &=& x^{k+2-\ell}\left[\begin{array}{cc}
x^{\ell-1}+\sum a^{\ell-1-i}x^i  & u_{\half(\ell+1)}\Lab{12}\\
u_{\half(\ell+1)}\Lab{21} & x-a\\
\end{array}\right]~~,
\end{eqnarray}
which satisfy $JE=EJ=x^{k+2}\bfone$ due to the constraint
$u_{\half(\ell+1)}\Lab{12}u_{\half(\ell+1)}\Lab{21}= -a^{\ell+1}$.

%%%%%%%%%%%%%%%%%%%%%%%%%%%%%%%
\subsection{$\cM_1\oplus 
\cM_{1}\rightarrow \cM_{2}\oplus \cM_{0}$ or
$\cM_{-1}\oplus \cM_{3}$}

%%%%%%%%%%%%%%%%%%%%%%%%%%%%%%%

This example demonstrates how different non-trivial composites can be produced
from one D-brane configuration, by appropriately tuning moduli.
We consider the system $\cM_1\oplus \cM_{1}$
with a tachyon condensate specified by:
$$
J^{\ell_1=1,\ell_2=1} = \left[\begin{array}{ccc}
x^2-u_1\Lab{11}x-u_2\Lab{11} & -u_{2}\Lab{12}-u_{1}\Lab{12}x\\
-u_{2}\Lab{21}-u_{1}\Lab{21}x & x^2-u_1\Lab{22}x-u_2\Lab{22}\\
\end{array}\right]~~.
$$
To analyze the moduli space, we compute: 
\beq
\det J:=\det J=x^4-(a+g)x^3+(ag-ce-h-b)x^2+(ah+bg-de-cf)x+bh-df~~,
\eeq
where we introduced the simplified notation:
\beq
u_1\Lab{11}=a~,~u_2\Lab{11}=b~,~u_1\Lab{12}=c~,~u_2\Lab{12}=d~,~u_1\Lab{21}=e~,~u_2\Lab{21}=f~,~
u_1\Lab{22}=g~,~u_2\Lab{22}=h~~.
\eeq
The factorization locus is given by $\det J(x)=x^4$, which gives the equations:
\beq
{\cal Z}~~:~~a+g=0~~,~~ag-ce=h+b~~,~~bh=df~~,~~ah+bg=de+cf~~.
\eeq
Hence ${\cal Z}$ is a four-dimensional affine
variety, namely a complete intersection in $\C^8$. 
Since the first equation is linear, we can use it to eliminate the variable $g$ in
terms of $a$:
\beq
g=-a~~.
\eeq
This reduces the remaining relations to:
\begin{eqnarray}
\label{sys}
h+b+a^2&=&-ce~~\nn\\
bh&=& df~~\\
cf+de&=& a(h-b)~~,\nn
\end{eqnarray}
which present ${\cal Z}$ as a complete intersection in $\C^7$. 
If $a\neq 0$, we can eliminate $b$ and $h$ by using the first and third
equation:
\begin{eqnarray}
b&=&-\frac{ace+a^3+cf+de}{2a}\nn\\
h&=&-\frac{ace+a^3-cf-de}{2a}\nn~~.
\end{eqnarray}
Then the second relation in (\ref{sys}) defines the following hypersurface in $\C^5$:
\beq
{\cal Z}_{fact,0}~~:~~4a^2df=a^2(ce+a^2)^2-(cf+de)^2~~.
\eeq
Thus ${\cal Z}\setminus (a)={\cal Z}_{fact,0}\setminus (a)$, where $(a)$ denotes the
divisor $a=0$. For $a=0$, the system (\ref{sys}) becomes:
\begin{eqnarray}
\label{sys0}
h+b&=&-ce\nn\\
bh&=&df\\
cf+de&=&0\nn~~.
\end{eqnarray}
The first two conditions are the Viete relations for the polynomial
$y^2+cey+df$; thus $h$ and $b$ are the two solutions of the equation in $y$:
\beq
y^2+cey+df=0~~.
\eeq
The remaining condition in
(\ref{sys0}) is the defining equation of the conifold (ODP) singularity 
in three dimensions. Hence the subvariety ${\cal Z}\cap (a)$ is a
branched double cover of the conifold singularity.

Along the factorization locus, the matrix $J$ takes the form:
\beq
J=\left[\begin{array}{cc}x^2-ax-b&-cx-d\\-ex-f&x^2+ax-h\end{array}\right]~~, 
\eeq
whose parameters are subject to (\ref{sys}). Let us first assume that $ace\neq 0$.
Then the greatest common denominator of all $1\times 1$ minors of $J$ is:
\beq
\label{G1}
G_1=\gcd(x^2-ax-b,x^2+ax-h,x+\frac{d}{c}, x+\frac{f}{e})=
\gcd(x^2-ax-b,x-\frac{h-b}{2a},x+\frac{d}{c}, x+\frac{f}{e})~~.
\eeq
In the second equality, we performed a linear combination with constant
coefficients of the first two polynomials. It is clear that $G_1$ is nontrivial
(i.e. differs from a nonzero constant) if and only if the following
conditions are satisfied: 
\beq
\label{fracrels}
\frac{h-b}{2a}=-\frac{d}{c}=-\frac{f}{e}=\alpha~~,  
\eeq
with $\alpha$ a complex constant subject to: 
\beq
\label{alpha_eq}
\alpha^2-a\alpha-b=0~~.
\eeq
Conditions (\ref{fracrels}) mean that all linear polynomials in the last form
of  (\ref{G1}) equal $x-\alpha$, while (\ref{alpha_eq}) is the requirement that $x-\alpha$
divides $x^2-ax-b$. When these relations hold, we have $G_1=x-\alpha$. 

\begin{figure}[hbtp]
\begin{center}
\scalebox{0.6}{\input{moduli3.pstex_t}}
\end{center}
     \caption{Realization of the processes $\cM_1\oplus \cM_1\longrightarrow
       \cM_0\oplus \cM_2$ and $\cM_1\oplus \cM_1\longrightarrow \cM_{-1}\oplus \cM_3$ in
       the moduli space. }
  \label{moduli3}
\end{figure}
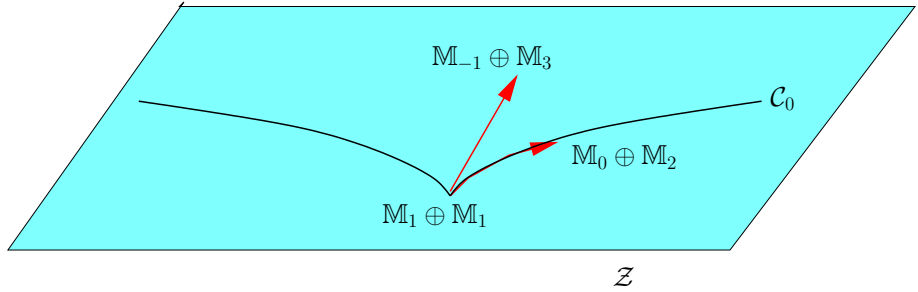

Equations (\ref{fracrels}) and (\ref{alpha_eq}) can be used to eliminate $b,d,f$ and $h$: 
\beq
\label{bdfh_elim}
b=\alpha^2-a\alpha, d=-\alpha c, f=-\alpha e, h=\alpha^2+a\alpha~~. 
\eeq
Substituting this into (\ref{sys}) gives the relations: 
\begin{eqnarray}
\alpha(a^2+ce)&=&0~~\nn\\
a^2+ce&=& -2\alpha^2~~\\
\alpha^2(\alpha^2-a^2-ce)&=&0~~,\nn
\end{eqnarray}
which are equivalent with $\alpha=a^2+ce=0$. Since $\alpha$ vanishes,
equations (\ref{bdfh_elim}) give $b=d=f=h=0$. Thus the locus ${\cal C}_0$ of nontrivial $G_1$ is 
$b=d=f=h=a^2+ce=0$, a subvariety of ${\cal Z}$ which is isomorphic with an
$A_1$ surface singularity. Along this
locus, one finds $G_1=x$ except at the origin, so that $J$ can be brought to the form:
\beq
J\sim \left[\begin{array}{cc}x&0\\0&x^3\end{array}\right]~~.
\eeq
This gives the minimal brane content
$\cM_{0}\oplus \cM_{2}$. By analyzing the case $ace=0$, one finds that 
the only exception occurs at the origin of the parameter space, which is 
the singular point of ${\cal C}_0$. There one finds $G_1=x^2$, with the
minimal brane content $\cM_{1}\oplus \cM_{1}$, which corresponds to the original
(undeformed) D-brane system. On the complement ${\cal Z}\setminus {\cal
  C}_0$, one has $G_1=1$ and $J\sim
\left[\begin{array}{cc}1&0\\0& x^4\end{array}\right]$, which gives the minimal brane
content $\cM_{-1}\oplus \cM_3$. Hence the factorization locus has the stratification:
\beq
{\cal Z}={\cal O}_{2,2}\sqcup {\cal O}_{1,3}\sqcup {\cal O}_{0,4}~~,
\eeq
where:
\begin{eqnarray}
{\cal O}_{2,2}&=&\{0\}~~{\rm (origin),~with~minimal~brane~content}~
\cM_1\oplus \cM_1\nn\\
{\cal O}_{1,3}&=&{\cal C}_0\setminus \{0\},~{\rm~with~minimal~brane~content}~
\cM_0\oplus \cM_2\nn\\
{\cal O}_{0,4}&=&{\cal Z}\setminus {\cal C}_0,{\rm~with~minimal~brane~content}~
\cM_{-1}\oplus \cM_3\nn~~.
\end{eqnarray}
Deforming from the origin into the stratum ${\cal O}_{1,3}$ implements the
process:
\beq
\cM_1\oplus \cM_1\longrightarrow \cM_0\oplus \cM_2~~,
\eeq 
while deformations into ${\cal O}_{0,4}$ lead to:
\beq
\cM_1\oplus \cM_1\longrightarrow \cM_{-1}\oplus \cM_3~~.
\eeq
This is shown schematically in figure \ref{moduli3}. 

\

Let us focus on the stratum  ${\cal O}_{1,3}$, 
noticing that its  normal space in $\C^8$ can parameterized by complex quantities
$\sigma_1, s\Lab{12}_2, s\Lab{21}_2,s_1\dots s_3$ defined through:
\begin{eqnarray}
u_1\Lab{11} &=& \sigma_1+\zeta_1,\nonumber\\
u_2\Lab{11} &=& -\sigma_1\zeta_1,\nonumber\\
u_1\Lab{22} &=& s_1-\zeta_1,\nonumber\\
u_2\Lab{22} &=& s_1\zeta_1,\nonumber\\
u_{1}\Lab{12}u_{1}\Lab{21} &=& s_2-{\zeta_1}^2\\
u_{1}\Lab{12}u_{2}\Lab{21} &=& s_3+s_1{\zeta_1}^2\nonumber\\
u_{2}\Lab{12}u_{1}\Lab{21} &=& -s_2\sigma_1+\sigma_1{\zeta_1}^2\nonumber\\
u_{2}\Lab{12}u_{2}\Lab{21} &=&  s_2\Lab{12} s_2\Lab{21}-s_3 \sigma_1-s_1\sigma_1{\zeta_1}^2~~.
\nonumber
\end{eqnarray}
Performing this change of variables brings the effective potential to the form: 
\beql{newW}
\weff^{\ell_1=1,\ell_2=1}\rightarrow 
\log\Big[(1-\sigma_1)(x^3-s_1 x^2-s_2 x-s_3)- s_2\Lab{12} s_2\Lab{21}\Big]
\Big|_{x^{-k-3}} \ =\ \weff(\sigma,s)^{\ell_1=0,\ell_2=2}.
\eeql
The variable $\zeta_1$ and the ratios of the off-diagonal boundary
changing moduli decouple, and represent flat directions
along which composite formation occurs.
Accordingly, matrix factorization persists along the
locus ${\cal C}_0$, as can be checked from the expressions:
\begin{eqnarray}
J^{\ell_1=0,\ell_2=2} ~~&=& \left[\begin{array}{ccc}
x^2-x\,{\zeta} & -x\,u_{1}\Lab{12}\\
-x\,u_{1}\Lab{21} & x^2+x\,{\zeta}\\
\end{array}\right]~~,
\\
E^{\ell_1=0,\ell_2=2} &=&
 x^{k-2}\left[\begin{array}{cc}
x^2+x{\zeta} & x\,u_{1}\Lab{12}\\
x\,u_{1}\Lab{21} & x^2-x{\zeta}\\
\end{array}\right]~~,
\end{eqnarray}
by using the constraints given above.

\section{Conclusions and outlook}

We studied moduli spaces and tachyon condensation for D-branes in
B-twisted minimal models of type $A_{k+1}$ and their massive deformations. In
particular, we showed that any D-brane in such models 
is isomorphic with a direct sum of `minimal' rank one
objects, which generalize the `rational' branes known from the conformal point. This
explains in what sense such branes play a distinguished role in minimal
models. It is important to realize that the isomorphism relating a D-brane
to a direct sum of minimal branes is {\em not} irrelevant and has
a nontrivial physical realization on the world-sheet. In particular, the parameters of
such isomorphisms are responsible for the fact that generic D-brane moduli
spaces are algebraic varieties of positive dimension (as opposed to
discrete collections of points). Therefore, it is {\em not} true that
minimal branes exhaust the collection of boundary sectors in such
models. On the contrary, the full D-brane category contains a
continuous infinity of objects, while the minimal
subcategory is finite. This distinction is physically meaningful and 
not a mathematical artifact.

We also showed that minimal brane
decompositions induce a stratification of each D-brane's moduli space,
where every stratum is associated with a different minimal brane content.
Varying moduli inside a given stratum amounts to changing the
isomorphism between the given D-brane and a fixed minimal brane decomposition,
while crossing from a stratum to another amounts to changing the
D-brane's minimal brane content. A combination of these processes implements transitions
between different systems of independent minimal branes. 
As in \cite{CIL2,CIL8}, our description is purely topological and
should be supplemented with a stability condition, whose proper
formulation in Landau-Ginzburg models is still unknown; we plan to 
return to this issue in future work.  

A central point of the present paper is our proposal of a closed, synthetic
expression for the effective tree-level potential $\weff$ of open
B-twisted strings ending on an arbitrary B-type brane. Since 
this quantity  is the generating functional of open string
scattering amplitudes, our generalized residue formula encodes the totality of such
amplitudes for the B-twisted string. At the conformal point, this
recovers all tree-level integrated CFT amplitudes containing only
chiral primary insertions. Our approach presents $\weff$ as a function 
of linearized deformation parameters, which play the physical role of
`special coordinates'. This goes beyond the mere computation of F-term
equations through algebraic homotopy theory, which is ambiguous due to
the freedom of choosing a minimal model of the associated differential
graded algebra, an ambiguity which amounts to performing {\em power series} redefinitions of 
coordinates along the {\em formal} moduli space \cite{CIL4}. The
geometric sense in which our coordinates are 'special' deserves
further investigation, and we plan to report on this issue in future
work. Another open issue under consideration is giving a direct derivation
of the tree-level potential $\weff$.

Our proposal for $\weff$ admits a holomorphic matrix model description, which makes contact
with the intuition that general branes in such
models can be described as collections of  D0-branes.
More precisely, $\weff$ arises as the {\em classical} potential of
such a matrix model. It is natural to conjecture that the
partition function of this model describes the coupling to topological
gravity on the world-sheet. Establishing such a conjecture along the
lines of \cite{KMM} requires a
detailed analysis of topological gravity on bordered Riemann
surfaces, which has not yet been performed. 

Another extension of the present work concerns 
D-branes in general  B-twisted Landau-Ginzburg models
\cite{Labastida}, as formulated in \cite{coupling, traces}. In the general case,
D-brane deformations correspond to the moduli space of certain
superconnections defined on the target space $X$ of the model (which
is a non-compact Calabi-Yau manifold), and one expects a much
more complicated description. However, the residue-like 
proposal for $\weff$ should generalize. One can expect 
substantial complications even for the simple case $X=\C^d$ with $d>1$, since
multivariate polynomial matrices do not generally admit a reduction to 
normal Smith form\footnote{Reduction to a matrix problem
  holds for $X=\C^n$ since finitely generated projective modules over $\C[x_1\dots x_n]$ are
  free by the Quillen-Suslin proof of Serre's conjecture.}.

\paragraph{\bf Acknowledgments}

The authors thank Ilka Brunner, Stefan Fredenhagen, 
Matthias Gaberdiel and Marcos Marino for stimulating conversations.
C.I.L thanks Albrecht Klemm for support and interest in his
work. W.L. and C.I.L. thank the Kavli Institute for Theoretical
Physics and University of California at Santa Barbara for a pleasant
stay, during which the present collaboration was formed.  This
research was supported in part by the National Science Foundation
under Grant No. PHY99-07949.

\appendix

%%%%%%%%%%%%%%%%%%%%%%%%%%%%%%%%%%%%%%%%%%%%%

\end{document}

%% file: cardy_moduli.pstex_t
\begin{picture}(0,0)%
\includegraphics{cardy_moduli.pstex}%
\end{picture}%
\setlength{\unitlength}{4144sp}%
\begingroup\makeatletter\ifx\SetFigFont\undefined%
\gdef\SetFigFont#1#2#3#4#5{%
  \reset@font\fontsize{#1}{#2pt}%
  \fontfamily{#3}\fontseries{#4}\fontshape{#5}%
  \selectfont}%
\fi\endgroup%
\begin{picture}(9990,6690)(496,-6946)
\put(10486,-6946){\makebox(0,0)[lb]{\smash{\SetFigFont{34}{40.8}{\rmdefault}{\bfdefault}{\updefault}{t}%
}}}
\put(496,-511){\makebox(0,0)[lb]{\smash{\SetFigFont{34}{40.8}{\rmdefault}{\bfdefault}{\updefault}{ u}%
}}}
\put(8371,-4246){\makebox(0,0)[lb]{\smash{\SetFigFont{29}{34.8}{\rmdefault}{\bfdefault}{\updefault}{C}%
}}}
\end{picture}

%% file: moduli1.pstex_t
\begin{picture}(0,0)%
\includegraphics{moduli1.pstex}%
\end{picture}%
\setlength{\unitlength}{4144sp}%
\begingroup\makeatletter\ifx\SetFigFont\undefined%
\gdef\SetFigFont#1#2#3#4#5{%
  \reset@font\fontsize{#1}{#2pt}%
  \fontfamily{#3}\fontseries{#4}\fontshape{#5}%
  \selectfont}%
\fi\endgroup%
\begin{picture}(6672,2248)(1294,-4817)
\put(7966,-3256){\makebox(0,0)[lb]{\smash{\SetFigFont{17}{20.4}{\rmdefault}{\bfdefault}{\updefault}{${\cal Z}$}%
}}}
\put(5671,-4741){\makebox(0,0)[lb]{\smash{\SetFigFont{17}{20.4}{\rmdefault}{\bfdefault}{\updefault}{$N_{\cal Z}$}%
}}}
\end{picture}

%% file: moduli3.pstex_t
\begin{picture}(0,0)%
\includegraphics{moduli3.pstex}%
\end{picture}%
\setlength{\unitlength}{4144sp}%
\begingroup\makeatletter\ifx\SetFigFont\undefined%
\gdef\SetFigFont#1#2#3#4#5{%
  \reset@font\fontsize{#1}{#2pt}%
  \fontfamily{#3}\fontseries{#4}\fontshape{#5}%
  \selectfont}%
\fi\endgroup%
\begin{picture}(9069,2928)(-11,-5092)
\put(7606,-3211){\makebox(0,0)[lb]{\smash{\SetFigFont{17}{20.4}{\familydefault}{\mddefault}{\updefault}{${\cal C}_0$}%
}}}
\put(6031,-5011){\makebox(0,0)[lb]{\smash{\SetFigFont{17}{20.4}{\rmdefault}{\bfdefault}{\updefault}{${\cal Z}$}%
}}}
\put(5626,-3796){\makebox(0,0)[lb]{\smash{\SetFigFont{17}{20.4}{\familydefault}{\mddefault}{\updefault}{$\cM_0\oplus \cM_2$}%
}}}
\put(4231,-2806){\makebox(0,0)[lb]{\smash{\SetFigFont{17}{20.4}{\familydefault}{\mddefault}{\updefault}{$\cM_{-1}\oplus \cM_3$}%
}}}
\put(3736,-4381){\makebox(0,0)[lb]{\smash{\SetFigFont{17}{20.4}{\familydefault}{\mddefault}{\updefault}{$\cM_1\oplus \cM_1$}%
}}}
\end{picture}